\documentclass[aps,amsmath,amssymb,reprint,superscriptaddress,floatfix]{revtex4-2}

\usepackage{bm}
\usepackage[normalem]{ulem}
\usepackage[utf8]{inputenc}
\usepackage[T1]{fontenc}
\usepackage{newtxtext,newtxmath}
\usepackage{hyperref}
\usepackage[capitalize]{cleveref}
\usepackage{appendix}
\usepackage{cancel}
\usepackage{todonotes}
\usepackage{physics}
\usepackage{siunitx}
\usepackage[caption=false]{subfig}
\usepackage[bbgreekl]{mathbbol}
\usepackage{color,soul}
\usepackage{graphicx}
\graphicspath{{figures/}}
\usepackage{booktabs}

\newcommand{\kb}{k_\mathrm{B}}
\newcommand{\kp}{k_\parallel}

\newcommand{\ct}{c_\mathrm{T}}
\newcommand{\cl}{c_\mathrm{L}}
\newcommand{\vecu}{\boldsymbol{u}}
\newcommand{\ca}{c_\alpha}
\newcommand{\qt}{q_\mathrm{T}}
\newcommand{\ql}{q_\mathrm{L}}

\newcommand{\ut}{u^\mathrm{T}}
\renewcommand{\ul}{u^\mathrm{L}}
\newcommand{\Fps}{F^{\mathrm{ps}}}

\newcommand{\vkp}{\bm{\kp}}

\newcommand{\zela}{\zeta^{\mathrm{ela}}}
\newcommand{\zvis}{\zeta^{\mathrm{vis}}}
\newcommand{\su}{\mathrm{[s.u.]}}
{\end{aligned}
\end{equation}}

\newcommand{\na}{n_\mathrm{A}}
\newcommand{\REMOVE}[1]{}

\newcommand{\avg}[1]{\left< #1 \right>}

\begin{document}

\preprint{APS/123-QED}

\title{Noncontact friction: Role of phonon damping and its nonuniversality}
\author{Miru Lee}
\email{miru.lee@uni-goettingen.de}
\affiliation{Institute for Theoretical Physics, Georg-August-Universit\"at 
G\"ottingen, 37073 G\"ottingen, Germany}
\author{Richard L.~C. Vink}
\affiliation{Institute of Materials Physics, Georg-August-Universit\"at 
G\"ottingen, 37073 G\"ottingen, Germany}
\author{Cynthia A. Volkert}
\affiliation{Institute of Materials Physics, Georg-August-Universit\"at 
G\"ottingen, 37073 G\"ottingen, Germany}
\author{Matthias Kr\"uger}
\email{matthias.kruger@uni-goettingen.de}
\affiliation{Institute for Theoretical Physics, Georg-August-Universit\"at 
G\"ottingen, 37073 G\"ottingen, Germany}

\date{\today}

\begin{abstract}
While obtaining theoretical predictions for dissipation during sliding motion is a difficult task, one regime that allows for analytical results is the so-called noncontact regime, where a probe is weakly interacting with the surface over which it moves. Studying this regime for a model crystal, we extend previously obtained analytical results and confirm them quantitatively via particle based computer simulations. Accessing the subtle regime of weak coupling in simulations is possible via use of Green-Kubo relations. The analysis allows to extract and compare the two paradigmatic mechanisms that have been found to lead to dissipation: phonon radiation, prevailing even in a purely elastic solid, and phonon damping, e.g., caused by viscous motion of crystal atoms. While phonon radiation is dominant at large probe-surface distances, phonon damping dominates at small distances. Phonon radiation is furthermore a pairwise additive phenomenon so that the dissipation due to interaction with different parts (areas) of the surface adds up. This additive scaling results from a general one-to-one mapping between the mean probe-surface force and the friction due to phonon radiation, irrespective of the nature of the underlying pairwise interaction. In contrast, phonon damping is strongly nonadditive, and no such general relation exists. We show that for certain cases, the dissipation can even {\it decrease} with increasing surface area the probe interacts with. The above properties, which are rooted in the spatial correlations of surface fluctuations, are expected to have important consequences when interpreting experimental measurements, as well as scaling with system size.

\end{abstract}

\maketitle

\section{Introduction}
Atomic force microscopy (AFM) provides a fascinating possibility to investigate the phenomenon of sliding friction on small length scales. An AFM tip sliding over a surface is known to perform so-called {\color{black}stick-slip} motion~\cite{prandtl28a,tomlinson29a,muser11a,gnecco00a,socoliuc04a,maier05a,liu15a_experiment,bennewitz01a_experiment}, most naturally understood from the famous Prandtl-Tomlinson model \cite{prandtl28a,tomlinson29a}. 
A related question concerns the energy dissipation channels in such a sliding 
process; contributions have been found from electrostatic 
interactions~\cite{kisiel11a_experiment,qi08a,liebsch99a}, electron excitation 
on the conduction 
band~\cite{kisiel11a_experiment,dorofeyev99a,stipe01a_experiment}, and phonon 
dynamics~\cite{barel10a,persson85a,persson99a,volokitin06a,kisiel11a_experiment,vink19a_simulation,schmidt20a,weber21a,afferrante19a,bugnicourt17a,sukhomlinov21a,prasad17a,kajita09a,glosli93a,kwon12a,hu20a}.
More specifically, energy transport by phonons
has been conjectured to be responsible for remarkable properties in friction, 
e.g., in polaronic conductors, where a drastic increase of friction near a 
phase transition was observed~\cite{weber21a,schmidt20a}, or in super 
conductors~\cite{kisiel11a_experiment}. It has long been an open question, 
however, 
precisely what properties of phonons give rise to dissipation in friction; 
here, recent work suggests the importance of phonon 
damping~\cite{schmidt20a,weber21a,vink19a_simulation,kwon12a,prasad17a,panizon18a}.

Concrete descriptions of how the mechanical work of the AFM tip is dissipated into the motion of atoms or electrons are difficult to obtain,
due to the many mechanisms involved, and additionally due to stick-slip behavior that occurs in sliding motion.
Such complexity is reduced in the case where the probe is a certain distance away from the surface, and interacts only weakly with it. Experimentally, this resembles the so-called {\it noncontact mode}~\cite{gotsmann01a_experiment,kantorovich01a_theory,kantorovich01b_theory,kantorovich02a_theory,kantorovich04a_simulation,kantorovich05a_simulation,volokitin06a,stipe01a_experiment,dorofeyev99a}, which, compared to the contact sliding mode, has at least two simplifications due to the weakness of the coupling between probe and surface; i) non-linear processes such as stick-slip motion are absent, and ii) the probe hardly affects the dynamics of the surface atoms, so that the latter can be treated to a good approximation as if the probe was not present.

In this manuscript, we theoretically study the described scenario of {\color{black}noncontact friction of an asperity free solid} in detail, analyzing which phonon properties determine friction, thereby extending analytical results from literature~\cite{volokitin06a}.
Starting from a Kelvin-Voigt model for a viscoelastic solid, we find the spatial dependence of position correlations, from which, via a Green-Kubo relation, the dissipation (friction) is found. This model describes damped phonons, with the damping, for example, originating from scattering of phonons with other phonons, defects, or electrons~\cite{michel15a,findley13a,landau86a,lee55a,volokitin06a}. {\color{black}It is amended by a stochastic (noise) term to describe thermal fluctuations~\cite{kantorovich08a_theory}.}

{\color{black} 
The analytic results from such a model enable to understand the two distinct mechanisms of dissipation by phonon. First is related to transport of energy by phonons through the solid, denoted phonon radiation in the following. Second, the energy of phonons dissipates into hidden degrees of freedom via, e.g., the scattering processes mentioned above. This mechanism we denote phonon damping.}

The two contributions are found to yield distinct behaviors of the resulting dissipation of probe motion. For probe motion parallel to the surface, the contribution by phonon radiation vanishes at small frequencies \cite{volokitin06a}, so that phonon damping is generally dominant. For motion perpendicular to the surface, both phonon radiation and phonon damping contribute. When the probe is close to the surface, the contribution by phonon damping dominates, and vice versa. The crossover length scale that separates the two regimes is typically given by $\sim \eta/\rho c'$ where $\eta$, $\rho$, and $c'$ denote the material viscosity, its mass density, and the real part of the speed of sound, respectively. Furthermore, the contribution from phonon radiation is universal and additive; using different types of probe-surface interactions yields the same (universal) relation between dissipation and mean force, which was given in Ref.~\cite{volokitin06a}. In other words, different atoms or areas on the surface contribute in an additive way to dissipation. In contrast, phonon damping yields a nonuniversal, or nonadditive contribution; no universal relation exists between force and dissipation, and different types of forces yield qualitatively different results. Also, atoms or areas on the surface contribute to dissipation in a nonadditive fashion. In some cases, interaction with a larger surface area may even yield a smaller frictional force.

The second part of the manuscript provides a quantitative test of these analytical predictions in an atomistic computer simulation, which, to our knowledge, has not been presented before. Here, we mimic as closely as possible the system studied analytically, which results in a discrete version of the mentioned model. {\color{black}The atomistic simulation} successfully deals with the subtleties of a well defined temperature at vanishing phonon damping, as well as finite size effects~\cite{vink19a_simulation,kantorovich08a_theory,kantorovich08b_theory,benassi10a_theory,benassi12a_theory}. We observe agreement with analytical results. Not only does this imply that the atomistic simulation well represents a true semi-infinite bulk system, but it also allows to distinguish between phonon radiation and phonon damping.

While a brief account of contributions from phonon damping is given in 
Ref.~\cite{volokitin06a}, it appears not much appreciated in literature; most 
experimental and theoretical works analyze their data in terms of pure phonon 
radiation~\cite{kisiel11a_experiment,liebsch99a,buldum99a,prasad17a,cui05a,gotsmann01a_experiment,guggisberg00a,stipe01a_experiment}.
We address this issue by providing estimates for the relative importance of 
the two contributions  for some experimental parameters.
Taking into account phonon damping may provide a better quantitative understanding of friction.

The manuscript  is organized as follows. In~\cref{sec:theory}, we start with modeling the dynamics of a viscoelastic solid using a stochastic Kelvin-Voigt theory at two different length scales: macroscopic and microscopic.
Introducing a probe near the solid surface, we make use of a Green-Kubo 
relation to derive the friction emerging from the dynamics of the solid in the 
noncontact (weakly coupled) regime. In~\cref{sec:results}, we find an analytic 
expression for the friction tensor, followed by a  quantitative comparison 
between the analytic and numerical results  in~\cref{sec:comp}. We then, 
in~\cref{sec:short_range}, study the effect of  interaction range on the 
friction tensor to illuminate the (non)universal behavior of the two 
contributions. The relative weight of these contributions for typical  
experimental parameters
is analyzed in~\cref{sec:exp}. \Cref{sec:conc} contains the summary and concluding remarks.

\section{System: Viscoelastic Solid}\label{sec:theory}
\subsection{System}
\begin{figure}
\centering
\includegraphics[width=0.9\columnwidth]{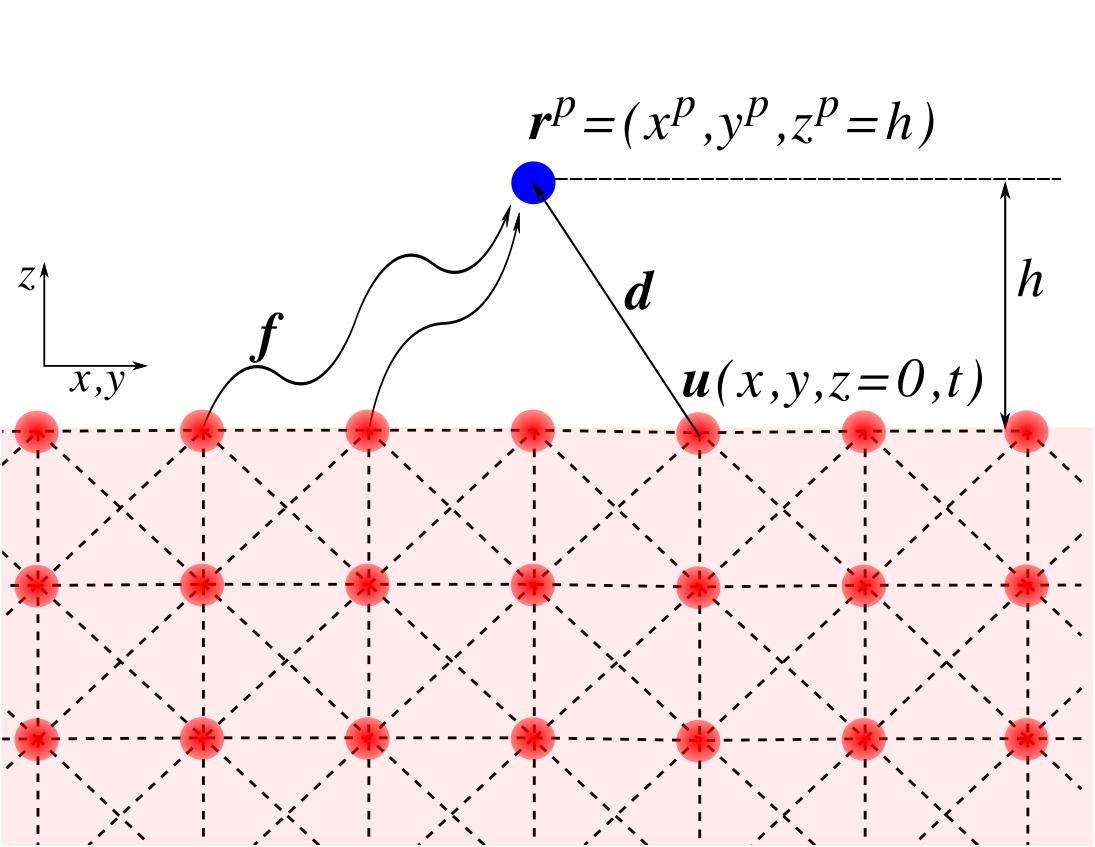}
\caption{Sketch of the investigated noncontact friction setup, with the 
blue particle being the (point) probe at $\bm{r}^p$. The square block 
represents the semi-infinite solid in the continuum description. The red 
particles are the crystal atoms in the discrete counterpart, where the 
dashed lines are the bonds among the nearest and the next nearest 
neighbors. The next nearest neighbor bonds render the shear modulus 
finite. The distance vector $\bm{d}$ denotes the distance between the probe 
and the lattice position of a surface atom with $h$ being the height of the 
probe, and $\bm{u}$ the fluctuating displacement vector field. These two 
vectors thus form a fluctuating distance vector, from which the 
instantaneous pairwise force $\bm{f}$ between the probe and a surface atom 
is evaluated.}
\label{fig:drawing}
\end{figure}

We aim to investigate the scenario depicted in~\cref{fig:drawing}, that is, a point probe at a height $h$ separated from a planar surface. The goal is to obtain the friction tensor of the probe in the limit where it is weakly coupled to the surface, i.e., to leading order in the probe surface interaction. We start from the thermal dynamics of the surface, from which, via a Green-Kubo relation, the friction tensor in the mentioned regime will be obtained.  

\subsection{Macroscopic continuum theory (analytical)}\label{sec:macro}

To model a semi-infinite viscoelastic solid, we begin with a Kelvin-Voigt model for a displacement (vector) field $u_i(\bm{r},t)$, a function of spatial position $\bm{r}$ and time $t$, and with vectorial index $i\in\{x,y,z\}$. It results from coarse-graining an (isotropic) crystalline structure, where each solid atom is connected to its neighbors by a dashpot harmonic spring~\cite{findley13a,landau86a}. 
In Fourier space, replacing time by frequency, $u_i(\bm{r},t) = \int_{-\infty}^{\infty}\frac{\dd{\omega}}{2\pi}u_i(\bm{r},\omega) e^{-i\omega t}$, the equation of motion reads \cite{findley13a,landau86a}
\begin{equation}
\omega^2\vecu(\bm{r},\omega)+\left(\cl^2-\ct^2\right)\nabla \nabla\cdot\vecu(\bm{r},\omega)+ \ct^2 \nabla^2\vecu(\bm{r},\omega)=0.
\label{eq:eqm}
\end{equation}
The speed of sound in the solid is generally complex (implying {\color{black}damping}), and a function of frequency. Since the simple-cubic lattice has a 1-atom basis, we need to distinguish between longitudinal and transverse acoustics waves. The respective speeds of sound are given by
\begin{equation}
\cl(\omega)=\sqrt{\frac{3K+4\mu-i\omega(3\xi+4\eta)}{3\rho}},~
\ct(\omega)=\sqrt{\frac{\mu-i\omega\eta}{\rho}}.
\label{eq:speed}
\end{equation}
Here, $K$ and $\mu$ ($\xi$ and $\eta$) are the bulk and shear elastic (viscous) moduli, respectively, and $\rho$ is the mass density.
{\color{black}Eq.~\eqref{eq:eqm}  describes sound waves, which are damped for finite viscous elements  $\xi$ and $\eta$, and the phonon decay length $\sim(\omega\Im[c^{-1}])^{-1}$ is finite. We thus expect that this equation describes a variety of systems where phonon damping may be caused, e.g., by phonon-phonon scattering or phonon-electron scattering.}

The linearity of Eq.~\eqref{eq:eqm} suggests introduction of a Green's tensor $G_{ij}(\bm{r},\bm{r}',\omega)$, giving the response of the displacement field $\bm{u}(\bm{r},\omega)$ upon perturbation via an external force field $\bm{f}^\mathrm{ext}(\bm{r},\omega)$,
\begin{equation}
u_i(\bm{r},\omega)=\int\dd[3]{r'}G_{ij}(\bm{r},\bm{r}',\omega) f^\mathrm{ext}_j(\bm{r}',\omega).\label{eq:G}
\end{equation}
The explicit expression of the Green's tensor for the semi-infinite solid, including the employed boundary conditions, can be found in~\cref{ap:GT_cal}~\cite{mindlin36a,landau86a,steketee58a,barbot10a,findley13a,lee55a,persson85a,persson01a,volokitin06a}.

While Eqs.~\eqref{eq:eqm} and \eqref{eq:G} are deterministic, one may consider thermal fluctuations to render the correlation function of the displacement field finite. Employing the fluctuation-dissipation theorem (FDT) for a system in equilibrium gives direct access to those via the Green's tensor~\cite{eckhardt84a,kruger11a,kruger12a,landau13a_stat_mech2,agarwal75a}, (note that $\avg{u_i(\bm{r})}=0$), 
\begin{equation}
\begin{aligned}
\avg{u_i(\bm{r})u_j(\bm{r}')}_\omega &:= \int_{-\infty}^{\infty}\frac{\dd{\omega'}}{2\pi}\avg{u_i(\bm{r},\omega)u_j(\bm{r}',\omega')}\\
&= \int_{-\infty}^{\infty}\dd{t}\avg{u_i(\bm{r},t)u_j(\bm{r}',0)}e^{i\omega t}\\
&= \frac{2\kb T}{\omega}\Im{G_{ij}(\bm{r},\bm{r}',\omega)},
\end{aligned}
\label{eq:FDT}
\end{equation}
where $\kb T$ is the temperature of the system, and the average brackets represent an ensemble average. \Cref{eq:FDT} with the explicit result for $G_{ij}$ is used in~\cref{sec:results} to obtain the friction of the probe, using the Green-Kubo relation introduced in Sec.~\ref{sec:wcl}.

\subsection{Microscopic theory (simulation)}\label{sec:micro}
{\color{black}In order to validate our analytic results obtained from the continuum theory, we compare them against molecular dynamics simulations below. In the simulations, we model the viscoelastic solid to numerically obtain~\cref{eq:FDT}.} We, therefore, introduce a discrete version of the viscoelastic solid. This amounts to filling the half space of Fig.\ref{fig:drawing} with a crystalline structure of atoms. {\color{black} We expect the discrete, microscopic description of the model solid to converge to the continuum, macroscopic counterpart defined in~\cref{eq:eqm} when the height of the probe is large compared to the lattice spacing (see~\cref{fig:drawing}).}

The equation of motion of the $m$th atom, where $\bm{U}_m(t)$ is its displacement from its lattice position (a vector), is given by Newton's second law,
\begin{equation}
M\ddot{\bm{U}}_m(t)=\sum_{n}\bm{F}^\mathrm{bond}_{mn}(t).
\label{eq:deqm}
\end{equation}
$M$ is the mass of the atom, and the summation over $n$ runs over the  (neighbor) atoms that are bonded to it.

The bonding force $\bm{F}^\mathrm{bond}_{mn}(t)$ consists of a (pairwise) elastic force, a damping force, and a random force,
\begin{equation}
\bm{F}^\mathrm{bond}_{mn}(t)=\bm{F}^\mathrm{spring}_{mn}(t)+\bm{F}^\mathrm{damp}_{mn}(t)+\bm{F}^\mathrm{ran}_{mn}(t),\quad \text{for}~m\ne n.
\label{eq:bond}
\end{equation}
The elastic force depends on the relative positions of the two atoms,
\begin{equation}
\bm{F}^{\mathrm{spring}}_{mn}(t)=-\kappa_{mn}\frac{\bm{L}_{mn}\bm{L}_{mn}}{\abs{\bm{L}_{mn}}^2}\cdot\left(\bm{U}_m(t)-\bm{U}_n(t)\right).\label{eq:Fs}
\end{equation}
The spring coefficients and the distance vector between the lattice positions of the two atoms are denoted by $\kappa_{mn}(=\kappa_{nm})$ and $\bm{L}_{mn}$, respectively. The form of Eq.~\eqref{eq:Fs} may look unfamiliar; it is the strictly linear version, prohibiting so-called geometric anharmonicity~\cite{norell16a}, which emerges even when using purely harmonic springs.
The damping force between two atoms, on the other hand, depends on their relative velocity,
\begin{equation}
\bm{F}^\mathrm{damp}_{mn}(t)=-\gamma_{mn}\frac{\bm{L}_{mn}\bm{L}_{mn}}{\abs{\bm{L}_{mn}}^2}\cdot\left(\dot{\bm{U}}_m(t)-\dot{\bm{U}}_n(t)\right)
\label{eq:dashpot}
\end{equation}
where $\gamma_{mn}(=\gamma_{nm})$ are the damping coefficients related to the two atoms.
It is also important to note that the damping force conserves momentum (locally), as does Eq.~\eqref{eq:eqm}. The form of the damping force resembles that of a dissipative particle dynamics (DPD)~\cite{moeendarbary09a}.

The random force $\bm{F}^\mathrm{ran}_{mn}$ has the following property, dictated by the fluctuation-dissipation theorem,
\begin{equation}
\begin{aligned}
\avg{\bm{F}^\mathrm{ran}_{mn}(t)}=&0\\
\avg{\bm{F}^\mathrm{ran}_{mn}(t)\bm{F}^\mathrm{ran}_{mn}(t')}=&2\kb T \gamma_{mn}\delta(t-t'),
\end{aligned}
\end{equation}
where the locality in time is due to the instantaneous form of the damping force. The numerical implementation of the random force is also very similar to a DPD simulation~\cite{moeendarbary09a}.

We construct a simple cubic crystal with lattice constant $a$. The atoms are connected to their nearest (NN) and next nearest neighbors (NNN) by the above mentioned bonds. For the simple cubic lattice, NNN bonds are needed to ensure a finite shear modulus. The spring (damping) coefficients for NN and NNN are denoted by $\kappa_1$ and $\kappa_2$ ($\gamma_1$ and $\gamma_2$), respectively.

The surface of the crystal is defined at $z=0$.
To mimic the semi-infinite solid, we impose the following boundary conditions. In the $xy$-directions, we use periodic boundary conditions. The bottom layer is frozen to prevent the crystal from moving vertically. The second bottom layer uses so-called stochastic boundary conditions~\cite{kantorovich08a_theory,kantorovich08b_theory,vink19a_simulation,benassi10a_theory,benassi12a_theory}. This serves two purposes in our numerical calculation. First, it reduces the vertical finite size effect of the system by partly absorbing the incoming phonons. Secondly, it helps to regulate the system's temperature when the system is nearly elastic. We also make sure that the system size is sufficiently large to avoid finite size effects. 

The displacement correlation can then be obtained by numerically integrating~\cref{eq:deqm} with the given boundary conditions in a standard molecular dynamics simulation~\cite{lammps}.

\subsection{Connection between continuum and discrete formulations}
In the limit of vanishing bond length, the simple cubic crystal reduces to the continuum isotropic viscoelastic solid. The coarse-graining can be formally done by expanding the dynamical matrix of the crystal around $\bm{k}=\bm{0}$ in $k$-space. Since the two theories model the same system, the speeds of sound in both descriptions must be identical. This allows us to directly relate the spring and damping coefficients to the elastic and viscous moduli (for the derivation see~\cref{ap:dispersion}),
\begin{equation}
\begin{aligned}
K=&\frac{1}{a}\left(\kappa_1+\frac{2}{3}\kappa_2\right),
\quad&\mu=\frac{1}{a}\kappa_2,\\
\xi=&\frac{1}{a}\left(\gamma_1+\frac{2}{3}\gamma_2\right),
\quad&\eta=\frac{1}{a}\gamma_2.
\label{eq:dcrel}
\end{aligned}
\end{equation}

\subsection{Friction of a weakly coupled point probe}\label{sec:wcl}
When moving with small amplitudes, the point particle probe with mass $m$ in Fig.~\ref{fig:drawing} at position ${\bm r}^p$ (with $z^p=h$) follows an equation of motion~\cite{zwanzig01a,persson85a,kantorovich08a_theory,ness15a}
\begin{align}
-\omega^2 r^p_i(\omega)m=-i \omega \Lambda_{ij}(h,\omega)r^p_j(\omega)+ F_i^{\rm ps}(h,\omega).
\label{eq:prob_eqm}
\end{align}
Here, $\Lambda_{ij}$ is the complex friction tensor, whose real part, $\Gamma_{ij}=\Re[\Lambda_{ij}]$ we aim to determine, and  $F_i^{\rm ps}(h,\omega)$ is the (stochastic) force acting between the probe and the surface. For the linear equation, Eq.~\eqref{eq:prob_eqm}, they are connected by  a Green-Kubo relation~\cite{kubo12a,zwanzig01a,kruger16a},
\begin{equation}
\Gamma_{ij}(h,\omega)=\frac{1}{2 \kb T}\int_{-\infty}^{\infty}\dd{t}\avg{\Fps_i(h,t) ; \Fps_j(h,0)}e^{i\omega t},
\label{eq:ff_cor_bare}
\end{equation}
where $\langle A;B\rangle=\langle (A-\langle A\rangle)(B-\langle B\rangle)\rangle$ denotes the covariance, which is finite in Eq.~\eqref{eq:ff_cor_bare} because of thermal fluctuations.
In the weak coupling limit, we evaluate the force covariance on the right hand side of Eq.~\eqref{eq:ff_cor_bare} under the dynamics in the \textit{absence} of the probe. The friction tensor is then naturally of second order of interaction forces, which is the leading order at large distance \cite{lee20a}. The dynamics of the semi-infinite solid in absence of the probe is obtained using the methods introduced in Secs.~\ref{sec:macro} and  \ref{sec:micro}.

{\color{black} A similar formalism for a generalized Langevin equation of a body coupled to another body can be found in Refs.~\cite{kantorovich08a_theory,ness15a};  a very general Langevin equation for a subset of degrees of freedom is found via the projection operator formalism \cite{zwanzig01a}.}

The atomistic simulation directly yields this force covariance, after specifying a pairwise force between atoms and probe (see~\cref{ap:simulation} for details).
This method provides a convenient way to obtain the friction tensor without introducing the probe in the simulated dynamics. Thus, in principle, one run of simulations determines any entry of $\Gamma_{ij}$ for any height $h$ and any pairwise type of force.  

The analytical derivation finds the force covariance from the fluctuations of the displacement field introduced in Sec.~\ref{sec:macro}, and allows to systematically extend the previously obtained friction tensor by~\citet{volokitin06a}.

We {\color{black} continue by assuming that the probe interacts with the surface particles via a pairwise potential} $V(\bm{s})$, whose negative gradient we denote the force $f_i$  (see~\cref{fig:drawing}). The instantaneous distance $\bm{s}$ between the probe ($x^p,y^p,z^p=h)$ and a surface atom or volume element $(x,y,z=0)$ fluctuates due to the fluctuations of the vector field $\bm u$, 
\begin{equation}\begin{split}
\bm{s}(x^p-x,y^p-y,h,t) = \bm{d}(x^p-x,y^p-y,h)+\bm{u}(x,y,t)
\label{eq:sto_dist}
\end{split}\end{equation}
with $\bm{d}(x-x^p,y-y^p,h)$ denoting the mean distance vector, since $\avg{\bm{u}}=0$.
The probe-sample force (for a given instance) reads, in the continuum model,
\begin{equation}
\Fps_i(h,t)=\iint_{-\infty}^{\infty}\dd{x}\dd{y}\na(x,y)f_i(\bm{s}),
\label{eq:fps}
\end{equation}
where the particle number per unit area $n_\mathrm{A}(x,y)$ is introduced to account the surface distribution of atoms~\cite{hamaker37a}. We assume it homogeneous, $n_\mathrm{A}(x,y)=n_\mathrm{A}$. {\color{black}In the discrete model, the probe-sample force is obtained by summing the pairwise force over surface atoms.}

Expanding the pairwise force for $|\bm{u}|\ll|\bm{d}|$, one gets
\begin{equation}
f_i(\bm{s})=f_{i}(\bm{d})
+(\partial_jf_{i}(\bm{s})|_{\bm{s}=\bm{d}})u_j(x,y,t) + \cdots.
\label{eq:fluc_pair}
\end{equation}
Notice that, for $\sqrt{\langle\bm{u}^2\rangle}\ll|\bm{d}|$, the first term is the average pairwise force and the second term is the fluctuating part. {\color{black} Also, in the limit of weak coupling between probe and solid considered here, the field $u_j$ in Eq.~\eqref{eq:fluc_pair} is independent of the probe position. This simplification can in principle be removed \cite{kantorovich08a_theory}.} We can thus exploit the Green-Kubo relation~\cref{eq:ff_cor_bare}, where the probe-sample covariance is calculated from the pairwise force {\color{black}(see Ref.~\cite{kantorovich08a_theory} for a similar expression for a discrete system)}
\begin{equation}
\begin{split}
\Gamma_{ij}(h,\omega)=&\frac{n_\mathrm{A}^2}{\omega}\Im\bigg\{\iiiint_{-\infty}^{\infty}\dd{x}\dd{y}\dd{x'}\dd{y'}\bigg.\\
&\bigg.\times
\partial_k f_{i}(\bm{d})G_{kl}(x,y,x',y',\omega)
\partial_l f_{j}(\bm{d}')\bigg\}
\end{split}
\label{eq:FDT_friction}
\end{equation}
with  notation $\partial_k f_i(\bm{d})\equiv\partial_k f_{i}(\bm{s})|_{\bm{s}=\bm{d}}$. Note that the force $f_i$ is real, and it can move inside the imaginary part. {\color{black}Note also that the integrals in~\cref{eq:FDT_friction,eq:fps} run over the surface of the solid because we have assumed that the probe interacts only with the atoms at the surface ($z=0$). The integration can be modified to include the depth coordinate, in which case the probe interacts with the atoms in lower layers as well. Future work will investigate how such assumptions are satisfied in experimental setups.}

The friction tensor $\Gamma_{ij}(h,\omega)$ is thus directly related to the imaginary part of the Green's function, via Eq.~\eqref{eq:FDT}, and the pairwise force.

We close the section with the following remark. The frequency $\omega$ in the friction tensor $\Gamma_{ij}(h,\omega)$, via, Eq.~\eqref{eq:prob_eqm}, carries a physical meaning in an AFM experiment; it is the oscillation frequency of the cantilever tip. This frequency  is typically of the order of $\SI{e5}{\hertz}$, which is small regarding the band structure of a typical solid.
We thus pay our attention to the behavior of the friction tensor~\cref{eq:FDT_friction} at small (vanishing) $\omega$.

\section{Analytical Results}\label{sec:results}
Let us  consider the case where the pairwise interaction potential between the probe and each surface atom is governed by an inverse power law of order $n$, $V(\bm{s})=\frac{\alpha}{|\bm{s}|^n}$, where $\alpha$ is an interaction coefficient of units of   length to the power $-n$ times energy. From $V$,  the pairwise force is,
\begin{equation}
f_i(\bm{s})=\frac{n\alpha s_i}{\abs{\bm{s}}^{n+2}}.
\label{eq:pair_int}
\end{equation}
The mean probe-surface force $\avg{\Fps_i(h)}$ then reads (for $n>1$, $\sqrt{\langle\bm{u}^2\rangle}\ll|\bm{d}|$)  
\begin{equation}\label{eq:mf}
\avg{\Fps_i(h)}=n_\mathrm{A}\iint_{-\infty}^\infty\dd{x}\dd{y}\avg{f_i(\bm{s})} = \frac{2\pi n_\mathrm{A}\alpha}{h^{n-1}}\delta_{iz}.
\end{equation}
Notice that mean probe-surface force points in the $z$-direction,  for symmetry. {\color{black} The force covariance is, however, finite for the parallel direction, so that a friction force in directions parallel to the surface arises.}

The friction tensor for the semi-infinite viscoelastic solid, expanded for small frequency $\omega$, is then found using Eq.~\eqref{eq:FDT_friction} , see~\cref{ap:friction_cal} for the details of the calculation. The entries of the tensor are given by
\begin{equation}
\begin{aligned}
\Gamma_{xx}(h,\omega)=&\Gamma_{yy}(h,\omega)\\=&\frac{1}{4\pi^2\rho\ct'^3}\left(\frac{\zela_{xx}(b)\omega^2}{\ct'^2}\avg{\Fps_z(h)}^2\right.\\
&+\eta\left.\frac{\zvis_{xx}(b,n)}{\rho\ct'h}\left(\dv{\avg{\Fps_z(h)}}{h}\right)^2 +\cdots\right)\\
\Gamma_{zz}(h,\omega)=&\frac{1}{4\pi^2\rho\ct'^3}\left(\dv{\avg{\Fps_z(h)}}{h}\right)^2\\
&\times\left(\zela_{zz}(b)+\eta\frac{\zvis_{zz}(b,n)}{\rho\ct'h}+\cdots\right)\\
\Gamma_{ij}(h,\omega)=& 0\qquad\qquad\text{if}~i\neq j
\label{eq:friction_results}
\end{aligned}
\end{equation}
with $\ct'=\Re{\ct(\omega)}=\sqrt{\frac{\mu}{\rho}}+\order{\omega^2}$. Here, $\zela_{ii}(b)$ and $\zvis_{ii}(b,n)$ are dimensionless prefactors depending on the ratio of speeds of sound, $b=\cl/\ct$, and the power $n$ in~\cref{eq:pair_int}. {\color{black}In obtaining~\cref{eq:friction_results}, we assumed that the ratio of speeds of sound $b$ is a real number for all $\omega$, which holds true if $\frac{K}{\mu}=\frac{\xi}{\mu}$.}

The first terms of $\Gamma_{ii}(h,\omega)$, corresponding to the contributions of phonon radiation, are independent of $n$, while the second terms, contributions from phonon damping, depend on $n$. This signifies a (non)universal behavior of the friction tensor which we discuss in more detail in~\cref{sec:short_range}.

Notice also that the friction tensor is diagonal, and $\Gamma_{xx}=\Gamma_{yy}$, due to the axial symmetry of the chosen crystal. From~\cref{eq:prob_eqm}, $\Gamma_{xx}$ is understood as the  parallel friction component and $\Gamma_{zz}$ as the perpendicular friction component.

\Cref{eq:friction_results} reduces to Eqs.~(35) and (36) of Ref.~\cite{volokitin06a} at $\eta=0$, i.e., the first terms of $\Gamma_{xx}$ and $\Gamma_{zz}$ (given in Ref.~\cite{volokitin06a} for $b=2$). 
The term linear in $\eta$ of $\Gamma_{xx}$ is equivalent to Eq.~(48) of Ref.~\cite{volokitin06a} for the case of a van der Waals interaction between a cylindrical probe and the surface. 

{\color{black}The friction tensor at $\eta=0$ accounts for the mechanism of excited sound waves that travel 
through the solid~\cite{persson85a,persson99a,hu20a} (are radiated away). 
We will refer to this contribution of phonon radiation as the elastic contribution since it prevails in a purely elastic solid.}

Notice that, at $\omega=0$, the parallel component $\Gamma_{xx}(h,0)$ {\color{black} is proportional to the viscosity $\eta$, while the perpendicular one $\Gamma_{zz}(h,0)$ exhibits also a term independent of $\eta$. This is due to the different waves excited by parallel and perpendicular motion.}

The contributions that depend on the viscosity $\eta$ result from the damped (or viscous) motion of solid atoms. We thus refer them as the viscous contribution. 
The viscous contribution survives at the limit of zero driving frequency.
This might be counter intuitive, as the phonon decay length diverges at 
$\omega\to 0$, and one may conclude that phonon damping is irrelevant. The 
viscosity $\eta$, nevertheless, plays an important role for the friction at 
$\omega\to 0$.

To conclude this section, we comment that our work goes beyond literature~\cite{volokitin06a} by i) inclusion of the viscous contribution for the $zz$ component, ii) extending the friction tensor for various powers of $n$,  by iii)  providing a quantitative analysis using atomistic simulations, and lastly by iv) a qualitative analysis of the (non)universality of the friction tensor.

\section{Theory and Simulation: Quantitative Comparison}\label{sec:comp}

\begin{figure}
\centering
\includegraphics[width=\columnwidth]{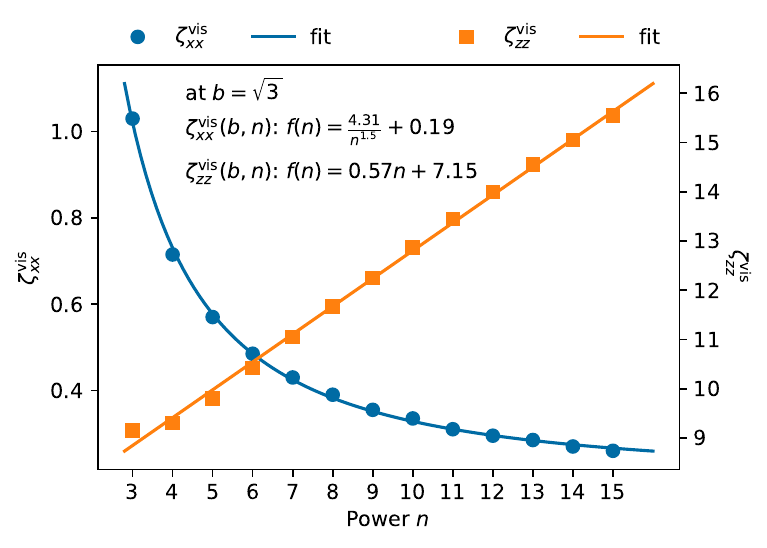}
\caption{The dimensionless prefactors of the viscous contribution, $\zvis_{xx}$ and $\zvis_{zz}$, as a function of power $n$ at $b=\sqrt{3}$. Solid lines represent fits, using the forms given in the figure.
}
\label{fig:zeta}
\end{figure}

\begin{figure*}
\centering
\subfloat{\centering
\includegraphics[width=.9\columnwidth]{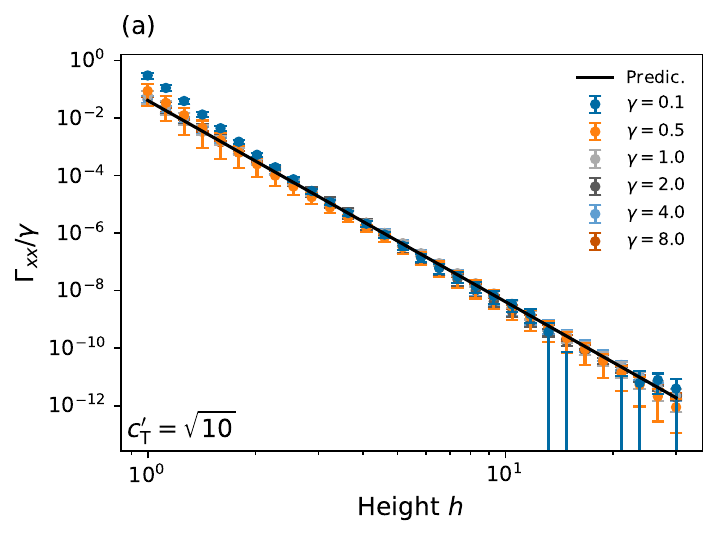}}
\subfloat{
\centering
\includegraphics[width=.9\columnwidth]{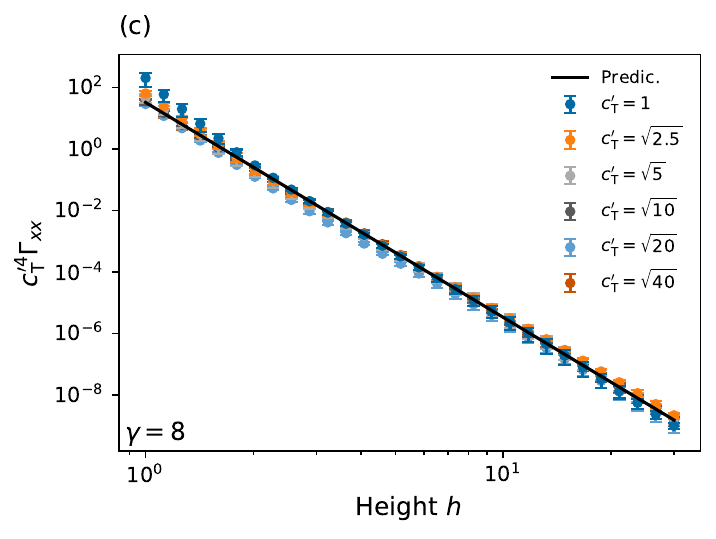}}\\
\centering
\subfloat{\centering
\includegraphics[width=.9\columnwidth]{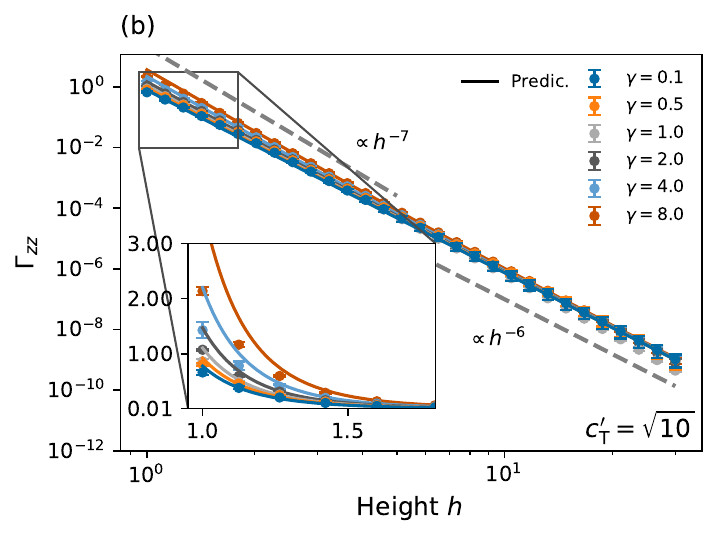}}
\subfloat{
\centering
\includegraphics[width=.9\columnwidth]{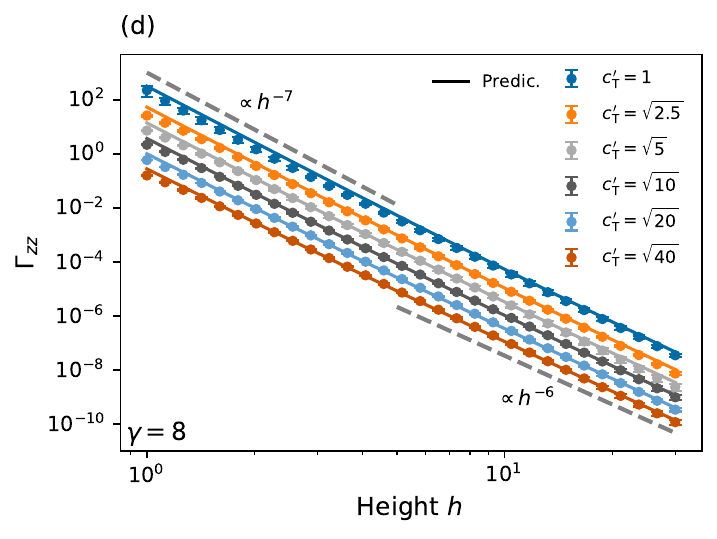}}
\caption{The friction tensor as a function of height $h$ at different material properties. {\color{black}In (a) and (b), the viscosity $\gamma$ is varied at fixed $\ct'=\sqrt{10}~\su$. In (c) and (d), the real part of the speed of sound is varied at fixed $\gamma=8~\su$.} The data points are the corresponding simulation results, in which we set $a=1$, $M=1$, $\alpha=1$ and $n_\mathrm{A}=1$. The height $h$ is given in the unit of lattice spacing $a$, and the system size is $(X,Y,Z)=(99a,99a,101a)$. The black lines in (a) and (c) denote the analytic prediction for the  master curve, respectively. The colored solid lines in (b) and (d) are the analytic predictions for the color matched material properties, and the gray dashed lines draw the inverse power laws in $h$. The inset in (b) shows a closer look at smaller $h$ with a linear scale on both axes, where the viscous contribution dominates. In Eq.~\eqref{eq:dfriction_results}, the crossover from the viscous to elastic contributions es expected at a height of  $h^*=1.56\gamma/\ct'$. The error bars, representing statistical errors, are estimated from the results found when using one fifth of the data.}
\label{fig:fric_comp}
\end{figure*}

\begin{figure*}
\subfloat{
\centering
\includegraphics[width=0.9\columnwidth]{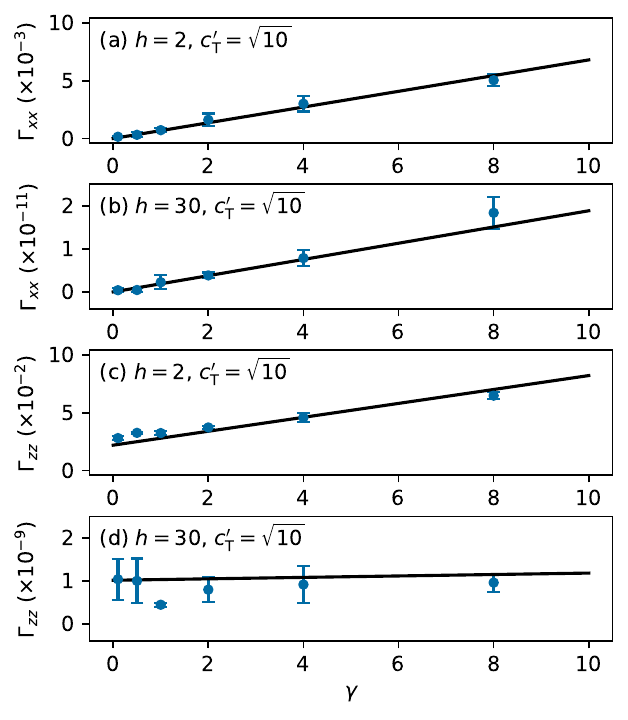}}
\subfloat{
\centering
\includegraphics[width=0.96\columnwidth]{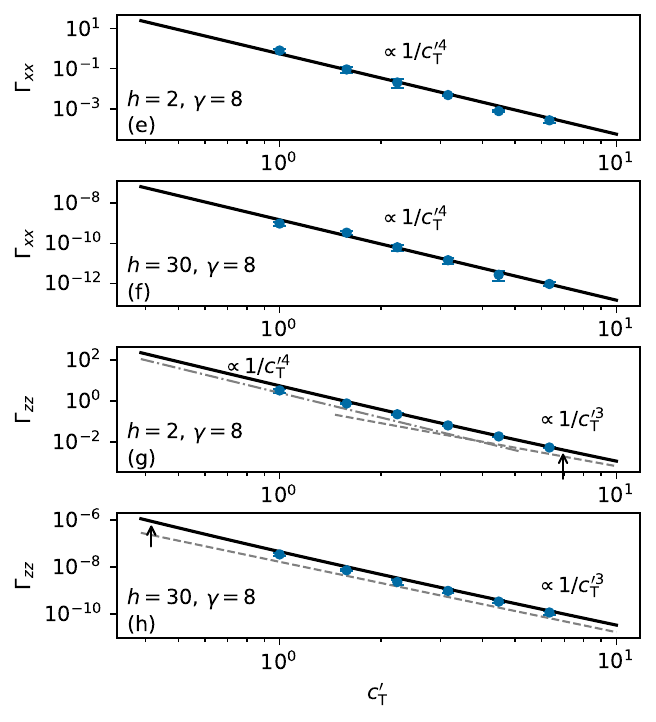}}
\caption{{\color{black}The friction tensor as a function of material properties at fixed heights $h$ in the unit of lattice spacing $a$. In the first column, we change the viscosity at fixed real part of the speed of sound $\ct'=\sqrt{10}~\su$, and in the second column, we change the real part of the speed of sound $\ct'$ at fixed viscosity $\gamma=8~\su$.} The data points denote the simulation results, and the black lines are the analytic predictions for each case. In (g) and (h) the gray dashed lines are the indicators for the power laws in $\ct'$, and the arrows mark the expected crossover from the viscous to the elastic contribution, given by $\ct^{\prime *}=1.58\gamma/h$. The statistical error bars, representing one standard deviation, are calculated by partitioning the time series of data into 5 pieces.
}
\label{fig:fric_gamma}
\end{figure*}

In this section, we quantitatively compare the analytic expressions in \cref{eq:friction_results} with the numerical data obtained in simulations, in the limit of $\omega\to0$.

To obtain numbers, we consider the case of $\kappa_1=\kappa_2\equiv\kappa$ and $\gamma_1=\gamma_2\equiv\gamma$, yielding $b=\sqrt{3}$ {\color{black}for all $\omega$} in Eq.~\eqref{eq:friction_results}, which is a realistic value in many materials, see~\cref{ap:cratio} for more details. From this choice, the dimensionless prefactors of the elastic contribution become $\zela_{xx} = 2.8$, and $\zela_{zz} = 5.9$. Those of viscous contribution, $\zvis_{xx}$ and $\zvis_{zz}$, are  functions of the power of the pairwise potential $n$ (see~\cref{fig:zeta}). These can be found for any value of $n$, and we have fitted phenomenological functional forms, for $n\ge3$,
\begin{equation}
\begin{aligned}
\zvis_{xx}(n)&\approx\frac{4.31}{n^{1.5}}+0.19\\
\zvis_{zz}(n)&\approx0.57n+7.15
\label{eq:zeta}
\end{aligned}
\end{equation}
Here we choose $n=3$ \footnote{The inverse power $n=3$ is a sufficient condition for~\cref{eq:FDT_friction} to converge, and is the smallest $n$ with which we can find an analytic expression of the friction tensor}, with $\zvis_{xx}=1.03$, and $\zvis_{zz}=9.15$.

Consequently, our simulation results are to be compared to the following predictions for the friction tensor,
\begin{equation}
\begin{aligned}
\Gamma_{xx}(h,0)=&\frac{4.12 n_\mathrm{A}^2\alpha^2\gamma}{a\rho^2\ct'^4h^7}\\
\Gamma_{zz}(h,0)=&\frac{n_\mathrm{A}^2\alpha^2}{\rho\ct'^3}\left(
\frac{23.4}{h^6} +\frac{36.5\gamma}{a\rho\ct'h^7}
\right),
\label{eq:dfriction_results}
\end{aligned}
\end{equation}
which thus apply in the limit $\omega\to 0$. Here, the viscosity $\eta$ is replaced by the damping coefficient $\gamma$ using the relation given in~\cref{eq:dcrel}. We thus use the terms viscosity and damping coefficient interchangeably. {\color{black} Also note that the units used in simulations measure length in terms of lattice spacing $a$, mass in terms of $M$ (mass of atoms), and time in units of  $\tau$, the Lennard-Jones time unit, which is derived from the energy unit (in our case from the system temperature $\kb T$, see~\cref{ap:simulation} for the details of the simulation units.). Simulation units are indicated by [s.u.].}

Let us first discuss the height dependence of the friction tensor~\cref{eq:dfriction_results} in relation to material properties. The parallel component of the friction tensor $\Gamma_{xx}$ is given by a single power law, i.e., here $\propto h^{-7}$, since it only picks up the viscous contribution due to phonon damping.
The perpendicular component $\Gamma_{zz}$, on the contrary, shows two distinct power laws. At a larger height $h$, it is dominated by the elastic contribution due to phonon radiation, signified by $h^{-6}$, whereas at a smaller height, it is so by the viscous contribution with $h^{-7}$. A crossover height is estimated to be $h^*=\frac{\zvis(b,n)}{2\zela(b)\rho}\frac{\eta}{\ct'}$ or at $h^*=1.56\frac{\gamma}{\ct'}$ in the simulations with the given parameters. 

This is demonstrated in~\cref{fig:fric_comp}(a) and (b), where we plot the friction tensor as a function of height $h$ for different values of the viscosity $\gamma$. Notice that the parallel friction component is normalized by $\gamma$, and the curves $\Gamma_{xx}/\gamma$ collapse into a master curve. This shows that $\Gamma_{xx}$ is a linear function of $\gamma$ as predicted by Eq.~\eqref{eq:dfriction_results}. Also quantitatively, the analytic prediction of the master curve agrees very well with our measurements, where we emphasize that there is no free parameter in this comparison.

There exists, however, a visible deviation for very large and very small values of $h$, which we attribute to finite size effects. {\color{black}For small $h(\approx a)$, the continuum description is not valid and the atomic structure felt by the probe may have additional effects on the friction~\cite{panizon18a}}. For large values of $h$, we approach the size of the system, that is $(X,Y,Z)=(99a,99a,101a)$. These deviations are especially pronounced for small $\gamma$, which seems to imply that finite size effects introduce elastic contributions in $\Gamma_{xx}$, so that, for these regimes, $\Gamma_{xx}/\gamma$ gets large for $\gamma\to 0$. Error bars in the figure are derived from statistical uncertainty and from general difficulties of identification of the value of the integral; numerical evaluation of the Green-Kubo integral~\cref{eq:ff_cor_bare} is difficult, as the integrand is plagued by finite size effects for large times $t$~\cite{benassi10a_theory,benassi12a_theory}, and the error grows for large times as $\sqrt{t}$~\cite{kiryl19a}. See~\cref{ap:simulation} for details on evaluating the Green-Kubo integral from simulation data.

As discussed above, because $\Gamma_{zz}$ contains both viscous and elastic 
contributions, the perpendicular component exhibits a crossover from viscous 
dominant to elastic dominant behaviors upon varying the height $h$. Even though 
finite size effects plague the limits of small and large $h$, a small 
window of $h$ is accessible, and this crossover can be displayed in the simulation 
in~\cref{fig:fric_comp}(b). For example, for $\gamma=8~\su$, the crossover height 
is expected at $h\approx 4~\su$ in the given units, and indeed, that curve crosses 
over from $h^{-6}$ to $h^{-7}$. The inset in~\cref{fig:fric_comp}(b) shows the 
small $h$ behavior in a linear scale, emphasizing that the curves strongly 
depend on $\gamma$ there. In the opposite limit of large $h$, the curves depend 
less on $\gamma$, as expected from Eq.~\eqref{eq:dfriction_results}.

Similar analyses are presented in~\cref{fig:fric_comp}(c) and (d), by varying the real part of the speed of sound, $\ct'$. When multiplied with $\ct'^4$, the parallel component of the friction tensor, $\Gamma_{xx}$, again yields a master curve, which is accurately predicted by the analytic expression. $\Gamma_{zz}$ also agrees with our prediction at the various $\ct'$. In this case, the crossover length moves to the left with increasing $\ct'$.

In~\cref{fig:fric_gamma}, we investigate the friction tensor upon varying the viscosity, $\eta$, and the real part of the speed of sound, $\ct'$, at fixed heights. The parallel component is proportional to $\gamma$ and $\ct'^{-4}$ regardless of the height $h$. The perpendicular component, on the other hand, changes its dependence on $\gamma$ and $\ct'$ depending on the height $h$. At $h=2~\su$, the ordinate in~\cref{fig:fric_gamma}(c), corresponding to the elastic contribution, is small compared to the linear growth with $\gamma$ for the range shown. This marks that the viscous contribution is considerable. The result in~\cref{fig:fric_gamma}(g) is thus given by a combination of $\ct'^{-4}$ and $\ct'^{-3}$ with the crossover marked in the graph.
At $h=30~\su$, $\Gamma_{zz}$ is dominated by the elastic contribution, as the ordinate in~\cref{fig:fric_gamma}(d) is large compared to the increase with $\gamma$, for the shown range. Additionally, the crossover from the viscous to elastic contribution happens at a small $\ct'$ as shown in~\cref{fig:fric_gamma}(h), leaving the friction, to a good approximation, with being proportional to $\ct'^{-3}$.

Summarizing, \cref{fig:fric_comp,fig:fric_gamma} quantitatively illustrate the fundamental difference between the perpendicular, $\Gamma_{zz}$, and the parallel, $\Gamma_{xx}$, components of the friction tensor. The perpendicular friction shows both of the elastic contribution (due to phonon radiation) as well as the viscous contribution (due to phonon damping). Whereas, the parallel friction only shows the viscous contribution. Furthermore, our simulation measurements in $\Gamma_{zz}$ demonstrate that, depending on the height of the probe and material properties, one contribution may dominate over the other.

\section{(non)Universality and (Non)Additivity of probe friction}\label{sec:short_range}

We noted in Eq.~\eqref{eq:friction_results} a  significant difference between the elastic and viscous contributions; the elastic contribution yields a universal relation between friction and mean force. This relation depends on material properties, including the ratio of speeds of sound, but holds for any probe-surface interaction (thus the term {\it universal}). Indeed, we find that the formula for the elastic contribution remains valid for any pairwise form of the interaction potential as long as the double surface integral in~\cref{eq:FDT_friction} remains finite. For the viscous contribution however, this is not the case, as, e.g., the power $n$ in the pairwise interaction enters nontrivially in   $\zvis_{ii}(b,n)$ in Eq.~\eqref{eq:friction_results}. 

\begin{figure*}
\subfloat{
\centering
\includegraphics[width=0.95\columnwidth]{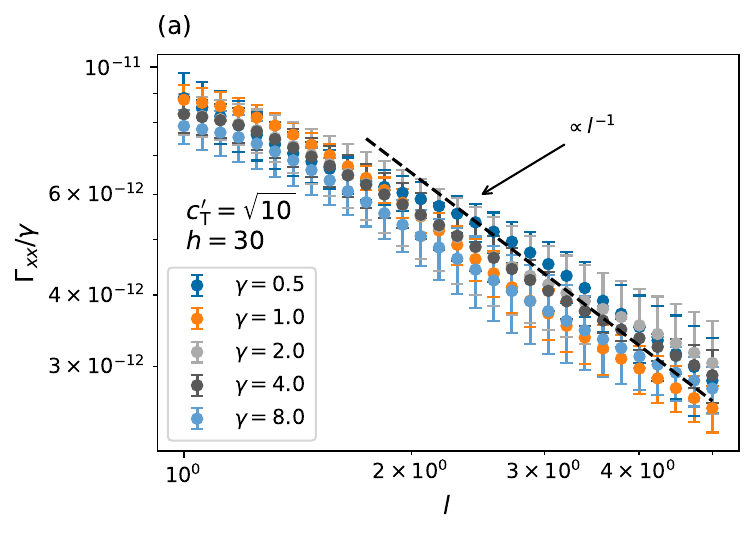}}
\subfloat{
\centering
\includegraphics[width=0.88\columnwidth]{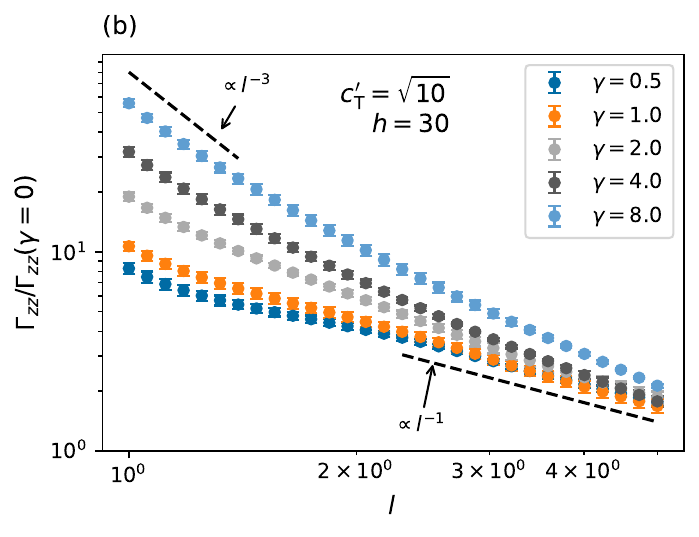}}
\caption{{\color{black}The interaction range dependence of the friction tensor, the parallel (a) and perpendicular (b) components measured in the simulation units. The interaction is $\Tilde{V}(x,y,h) = \alpha\exp{-\frac{x^2+y^2}{l^2}}(x^2+y^2+h^2)^{-3/2}.$ The probe height is fixed at $h=30$ in the unit of lattice spacing $a$}. The parallel $\Gamma_{xx}$ and perpendicular $\Gamma_{zz}$ components are normalized by the viscosity $\gamma$ and by the elastic contribution $\Gamma_{zz}(\gamma=0)$, respectively. The black dash lines indicate various power laws. The statistical error bars, representing one standard deviation, are estimated by partitioning the time series of data into 5 pieces.}
\label{fig:range}
\end{figure*}

This statement can be further illustrated by regarding a finite surface area the probe interacts with. This is achieved by using the 
following potential,
\begin{equation}
V(\bm{s}) = \exp{-\frac{s_x^2+s_y^2}{l^2}} \Tilde{V}(|\bm{s}|),
\label{eq:general_int}
\end{equation}
where $l$ plays the role of a lateral  interaction range, and $\Tilde{V}(|\bm{s}|)$ depends on $|\bm{s}|$ as denoted. 
For $l/h\gg1$, the case of~\cref{eq:pair_int} is recovered. On the contrary, if $l/h\ll1$, the interaction range is so localized that the dependence on $s_x$ and $s_y$ of the pairwise potential $\Tilde{V}(|\bm{s}|)$ becomes irrelevant,
\begin{equation}
\lim_{\frac{l}{h}\to0} V(\bm{s})= \exp{-\frac{s_x^2+s_y^2}{l^2}} \Tilde{V}(h).
\label{eq:short_int}
\end{equation}
This limit can be studied analytically. The mean force acting between probe and surface reads then (it has only a $z$-component),
\begin{equation}
\avg{\Fps(h)}=-\iint_{-\infty}^{\infty}\dd{x}\dd{y}n_\mathrm{A}\dv{\avg{V(\bm{s})}}{h}=-l^2 n_\mathrm{A} \pi \Tilde{V}'(h)
\end{equation}
with $\Tilde{V}'(h) = \dv*{\Tilde{V}(h)}{h}$.
In the discrete case, the number of atoms interacting with the probe is thus roughly given by $l^2/a^2$.

In this given limit, the friction tensor takes the insightful form (see~\cref{ap:local} for details, dots represent higher orders in $\omega$)
\begin{equation}
\begin{aligned}
\Gamma_{xx}(h,\omega)=&\frac{\zela_{xx}(b)l^4\omega^2\na^2{\Tilde{V}}'^2(h)}{4\ct'^5\rho}\\
+&\frac{\pi n_\mathrm{A}^2 \eta}{32l\rho^2\ct'^4}
\left(9\sqrt{2\pi}{\Tilde{V}}^2(h)
+8l \Tilde{V}(h)\Tilde{V}'(h)
\right.\\
&+\left.3l^2\sqrt{2\pi}\Tilde{V}'^2(h) \right)+\cdots\\
\Gamma_{zz}(h,\omega)=&\frac{\zela_{zz}(b)l^4 n_\mathrm{A}^2\Tilde{V}''^2(h)}{4\rho\ct'^3}\\
+&\frac{\pi l n_\mathrm{A}^2\eta}{16\ct'^4\rho^2}
\left(3\sqrt{2\pi}\Tilde{V}'^2(h)
+4l \Tilde{V}'(h)\Tilde{V}''(h)
\right.\\
&+\left. 3l^2\sqrt{2\pi}\Tilde{V}''^2(h)\right)+\cdots.
\label{eq:friction_range}
\end{aligned}
\end{equation}
\Cref{eq:friction_range} holds true for any $\Tilde{V}(h)$ at the given ratio of the speeds of sound, $b=\sqrt{3}$. Notably, when rewriting the elastic contributions of $\Gamma_{xx}(h,\omega)$ and $\Gamma_{zz}(h,\omega)$ of~\cref{eq:friction_range} in terms of $\avg{\Fps(h)}$ and $\dv*{\avg{\Fps(h)}}{h}$, respectively, one finds that the expression becomes identical to that of the long range interaction,~\cref{eq:friction_results}. This reiterates the above statements that this contribution, stemmed from phonon radiation, is universal and the relation between mean force and friction tensor is insensitive to the range of the interaction. The contribution naturally, for small $l$, grows as $l^4$, as each power of mean force grows as $l^2$. This also illustrates the statement of additivity; each surface element contributes in an additive way to the friction tensor. (Note that  the notion of additivity implies that different parts of the surface add independently to $\Gamma_{ii}$.  $\Gamma_{ii}$ is however not proportional to surface area.)  

The viscous contribution is different in that regard. First, the results shown in Eq.~\eqref{eq:friction_range} cannot be deduced from Eq.~\eqref{eq:friction_results}, pointing to nonuniversality of this case. Starting with the $zz$ component, we note that, for small $l$, the contribution from phonon damping is linear in $l$. This behavior is highly nonadditive, which means that, with decreasing $l$, the relative contribution of each surface area element {\it increases}. This effect is even more drastic for the $xx$ component; here the friction diverges with $l^{-1}$ for small $l$ {\color{black} despite the fact that the mean force between probe and surface vanish as $l$ goes to zero.}

Can these predictions be confirmed in the simulations? Not quantitatively, because, for small $l(\approx a)$, finite size effects are notable. For a qualitative check, we use
\begin{align}
V(\bm{s}) = \alpha\exp{-\frac{s_x^2+s_y^2}{l^2}}\abs{\bm{s}}^{-3},
\end{align}
and evaluate the potential at $h=30~\su$ in order to ensure $h\gg l$ so that~\cref{eq:friction_range} applies. 

The surprising statement that the friction, for $xx$, {\it grows} with decreasing $l$ is indeed seen also in the simulations, see~\cref{fig:range} a). The divergence with $1/l$ is naturally cut off in simulations, by the finite lattice spacing. It is, however, remarkable that a smaller amount of surface area the probe interacts with yields a larger value of friction.
For the perpendicular component, we also observe the nonadditive behavior in the simulations, where we see qualitative agreement with the scalings predicted by Eq.~\eqref{eq:friction_range}.

To better understand this (non)universal behavior, let us consider the Green's tensor of an infinite solid. In real space, (the imaginary part of) the Green's tensor of the transverse mode reads, with $z=z'$,
\begin{equation}
\begin{aligned}
&\Im{G^{\rm trans}_{ij}(x,y,x',y',\omega)}=\frac{\omega}{2\kb T}\avg{u_i(x,y)u_j(x',y')}_\omega\\
&\qquad=\frac{\delta_{ij}\omega}{4\pi\rho\ct'^3}\left(1+\frac{\eta}{\rho\ct'}\frac{1}{\sqrt{(x-x')^2+(y-y')^2}}+\cdots\right).
\end{aligned}\label{eq:corr}
\end{equation}
In calculating the friction tensor using~\cref{eq:FDT_friction}, the Green's tensor works as a kernel of the double surface integration of (the derivative of) the forces. Because the first term, which gives rise to the elastic contribution, is independent of the distance between the considered points, it does not have any effect on the integrand other than an overall multiplication. The elastic contribution thus results from integrating the forces, and universality and additivity follow. 

On the contrary, since the second term of the Green's tensor depends on $x-x'$ and $y-y'$, the resulting friction tensor cannot be generalized for an arbitrary functional form of force. It leaves us with the non-trivial and case specific scaling behavior of the viscous contribution in terms of the type of interaction.

This observation can translate to modal decomposition in wavevector space, here 
in the space of wavevectors $\kp$, parallel to the surface. At $\omega\to0$, 
the elastic contribution in the correlation function is $\sim\delta(\kp)$, and 
it  picks the mode with $\kp=0$ of the force. This $\kp=0$-mode corresponds to 
the total force between probe and surface, $\Fps_i=f_i(\kp=0)$, so that the 
above statements are reproduced. For the viscous contribution, the correlation 
function is  $\sim\kp^{-1}$, so that all modes contribute with corresponding 
weight.

Knowing the details of the interaction between the probe and the solid is imperative in understanding the friction, and the type of interaction may even allow for tunability, as exemplified in this section.

The dependence on $l$ in Eq.~\eqref{eq:friction_range} hints on interesting scaling behavior of the different contributions with probe size (interpreting $l$ as probe size), which we plan to investigate in future work~\cite{persson99a}.

\section{Discussion of typical experimental parameters}\label{sec:exp}
An important question is how phonon radiation and phonon damping are expected to contribute in an AFM setup. From~\cref{eq:friction_results} one can find expressions for the crossover heights between viscous and elastic contributions,
\begin{equation}
\begin{aligned}
h^*&\sim\left(\frac{\eta\ct'}{\omega^2\rho}\right)^{1/3}\qquad&\text{for}~\Gamma_{xx}\\
h^*&\sim\frac{\eta}{\rho\ct'}\qquad&\text{for}~\Gamma_{zz}.
\end{aligned}
\label{eq:transition}
\end{equation}
The parameters entering are the density $\rho$, the real part of speed of sound $\ct'$, the viscosity $\eta$, and the oscillation frequency $\omega$.

The driving frequency is in the order of $\SI{100}{\kilo\hertz}$ in a typical AFM experiment. The density and the transverse speed of sound of metals are of the orders of $\SI{e4}{\kilo\gram\per\meter^3}$ and $\SI{e3}{\meter\per\second}$, respectively. The viscosity of solids is a quantity which is not easily accessible, but it may be estimated to range between $\SIrange[range-phrase=\dots]{0.1}{1}{\pascal\second}$ for copper (at \SI{2}{\kilo\hertz}) and steel (at \SI{5}{\mega\hertz})~\cite{ono20a}, and $\SI{0.01}{\pascal\second}$ for aluminum (at \SI{50}{\giga\hertz})~\cite{bryner10a}. Notice that we infer the estimated values of viscosity from the measures of sound attenuation since directly measuring viscosity of metals below the melting temperatures is very difficult.

For the parallel friction $\Gamma_{xx}$, using these values, we estimate $h^*\sim\SIrange[range-phrase=\dots]{e-1}{1}{\milli\meter}$, above which the elastic contribution due to phonon radiation becomes as significant as the viscous counterpart. {\color{black}This implies that the elastic part, as found from this calculation, is likely irrelevant in a non-contact AFM measurement, since this height is astronomical from the  viewpoint of an AFM.}

For the perpendicular friction $\Gamma_{zz}$, the crossover is in a range of 
$\SIrange[range-phrase=\dots]{1}{100}{\nano\meter}$ {\color{black}, which is a typical AFM height range.} {\color{black} While our estimates for the viscosity include some uncertainty, our findings suggest i) that the viscous contribution can be dominant depending on the specific setup, and ii) that therefore the viscosity of the sample has to be considered in order to properly interpret a noncontact friction measurement.}

\section{Conclusion}\label{sec:conc}
In this paper, we study the noncontact friction of a semi-infinite viscoelastic solid. In modeling such a viscoelastic solid, we use a Kelvin-Voigt model, from which we find an analytic expression of the friction tensor, extending previous results \cite{volokitin06a}. We also construct a crystal system in MD simulations, modeling the same viscoelastic solid, and numerically calculate the friction tensor.

The noncontact friction consists of two distinct contributions: phonon radiation and phonon damping. The friction due to phonon radiation emerges due to the following mechanism; motion of the probe creates a phonon (or wave), which then propagates away, transporting energy into the infinitely large solid.  This mechanism prevails if the solid is purely elastic. We thus refer it to as the elastic contribution. As the propagating phonon is scattered with other phonons, defects, or electrons,  an additional contribution to the friction arises. As such damping behavior can be represented  by a viscosity, {\color{black}and} we call it the viscous contribution.

When the probe is moving perpendicular to the surface, the friction is given by the sum of both contributions. When the probe is moving parallel to the surface, however, it is only the viscous contribution that determines the friction, in the given ideal model. We demonstrate that the elastic and viscous contributions distinguish themselves with different inverse power laws in the height of the probe. This naturally gives rise to a crossover in the perpendicular friction, when one contribution overtakes the other. At small probe height, the viscous contribution dominates, whereas, at  higher height, it is the elastic contribution.

Our theoretical predictions are in quantitative agreement with the numerical calculations. This implies that one can precisely control the parallel and perpendicular frictions, given that one can manipulate the material properties.

We also illustrate a fundamental difference between phonon radiation and phonon 
damping mechanisms. At small frequency, the elastic contribution (due to phonon 
radiation) is universal, so that, for any microscopic interaction law acting 
between the probe and solid atoms, a unique relation between mean force and 
friction is found. As a consequence, the friction from this contribution is 
additive, i.e., different parts on the surface contribute independently. This 
behavior is rooted in the spatial correlations of the solid material.

The viscous contribution (due to phonon damping), on the contrary, depends non-trivially on the range and type of surface-probe interaction, due to the properties of spatial correlations. This can be exemplified by introduction of a  lateral interaction range. Especially for small interaction range, the nonadditivity of the viscous contribution is pronounced; for the $xx$ case, a smaller interaction range can even yield a larger friction, {\color{black} despite the fact that the mean force between probe and surface grows with interaction range. This result demonstrates that  mean force and friction are not necessarily related in non-contact mode.}

Future work will investigate the cases of anharmonic crystals as well as the question of scaling of these effects with probe size. {\color{black}Investigation of the validity of the linear Kelvin-Voigt model is also important, especially when quantum processes are dominant. It will also be important to study in more detail the parallel friction component in the presence of atomic surface structure, as in Ref.~\cite{panizon18a}.}

\acknowledgments
This work was funded by the Deutsche Forschungsgemeinschaft (DFG, German 
Research Foundation) 217133147/SFB 1073, project A01.

\appendix
\section{Green's tensor}\label{ap:GT_cal}
Solving~\cref{eq:eqm} for a semi-infinite solid is a variant of Boussinesq's problem and has been done with various methods~\cite{mindlin36a,landau86a,steketee58a,barbot10a,findley13a,lee55a,persson85a,persson99a,persson01a,volokitin06a}.

We define that the system has the surface on the $xy$-plane at $z=0$, and extends $-\infty<z\le0$. Let us rewrite the equation of motion~\cref{eq:eqm} so that it is better suited for finding the solutions
\begin{equation}
\left[\omega^2+\ca^2(\omega)\lambda^2\right]u_i^\alpha(z,\vkp,\omega)=0
\label{eq:ceqm2}
\end{equation}
with
\begin{equation}
\begin{aligned}
\bm{\lambda}=&i\vkp+\partial_z\hat{z},\\
\vkp=&k_x\hat{x}+k_y\hat{y},\\
\lambda=&\abs{\bm{\lambda}},\\
\kp=&\abs{\vkp}.
\end{aligned}
\end{equation}
The Fourier transform is defined as
$u_i(\bm{r},t)=\frac{1}{(2\pi)^3}\int_{-\infty}^{\infty}\dd^2{\bm{\kp}}\int_{-\infty}^\infty\dd{\omega}u_i(z,\bm{\kp},\omega) e^{i\left(k_x x+k_y y-\omega t\right)}$.

The boundary conditions are given by the stress tensor on the surface
\begin{equation}
\sigma_{ij}(z=0)=f^\mathrm{ext}_i\delta_{jz},
\label{eq:st_bound}
\end{equation}
where stress tensor is defined as
\begin{equation}
\sigma_{ij}(z,\vkp,\omega) = \rho\left(\cl^2-2\ct^2\right)\lambda_k u_k \delta_{ij}
+\rho\ct^2\left(\lambda_i u_j+\lambda_j 
u_i\right).
\end{equation}

To take advantage of the properties of the longitudinal and transverse modes, let us rewrite the displacement fields~\cite{arfken99a},
\begin{equation}\begin{aligned}
\ul_i(z,\vkp,\omega) &= \lambda_i \Phi(z,\vkp,\omega),\\
\ut_i(z,\vkp,\omega) &= \epsilon_{ijk}\lambda_j A_k(z,\vkp,\omega).
\end{aligned}\end{equation}
As a result,~\cref{eq:ceqm2} can be expressed as
\begin{equation}\begin{aligned}
\partial_z^2\Phi(z,\vkp,\omega) &= \ql^{2}\Phi(z,\vkp,\omega)\\
\partial_z^2A_{i}(z,\vkp,\omega)&= \qt^{2}A_{i}(z,\vkp,\omega),
\label{eq:eqm_phe}
\end{aligned}\end{equation}
where
\begin{equation}
\begin{aligned}
\qt^2(\vkp,\omega)&\equiv\kp^2 - 
\frac{\omega^2}{\abs{\ct(\omega)}^2} e^{i\psi(\omega)},\\
\ql^2(\vkp,\omega)&\equiv\kp^2 - 
\frac{\ct^2(\omega)}{\cl^2(\omega)}\frac{\omega^2}{\abs{\ct(\omega)}^2} 
e^{i\psi(\omega)}
\end{aligned}
\end{equation}
with
\begin{equation}\begin{aligned}
&\psi(\omega)=\atan{\left(\frac{2\ct'\ct''}{
	\ct'^2-\ct''^2}\right)},\\
&\ct'=\Re{\ct},\\
&\ct''=-\Im{\ct}.
\end{aligned}\end{equation}
Notice that $\ql(\vkp,\omega)$ and $\qt(\vkp,\omega)$ signify the branch points of $\kp$, at which they become zero (thus the dispersion relations),
\begin{equation}\begin{aligned}
\ql(\kp,\omega)=0\quad\text{at}~\kp=&\pm k_{b1}=\pm\frac{\ct}{\cl}\frac{\omega}{\abs{\ct}} e^{i\psi/2},\\
\qt(\kp,\omega)=0\quad\text{at}~\kp=&\pm k_{b2}=\pm \frac{\omega}{\abs{\ct}} e^{i\psi/2}.
\end{aligned}\end{equation}
These four points reside on the complex plane of $\kp$ since the speeds of sound are complex valued. How far those branch points are from the real axis is determined by $\psi(\omega)/2$, approximately the ratio of the real and imaginary part of the speed of sound $\ct(\omega)$, $\ct'' / \ct'$. In a purely elastic solid where $\psi(\omega)=0$ is true for all $\omega$, one expects the branch points are on the real axis of the $\kp$ complex plane, which is consistent with Refs.~\cite{persson85a,persson99a,volokitin06a}.

Trial solutions to the second order differential equations~\cref{eq:eqm_phe} are
\begin{equation}
\begin{aligned}
\Phi(z,\bm{k_\parallel},\omega) =& \Phi_0(\bm{k_\parallel},\omega) 
e^{\ql z},\\
A_i(z,\bm{k_\parallel},\omega) =& A_{i0}(\bm{k_\parallel},\omega) 
e^{\qt z}.
\end{aligned}
\end{equation}
Let us apply the boundary conditions~\cref{eq:st_bound}.
The first case is when the external force is exerted perpendicular to the surface,
\begin{equation}
\sigma_{iz}(z=0)=\begin{cases}
f^\mathrm{ext}_z\qquad&  \text{if}~i=z\\
0\qquad&  \text{otherwise}.
\end{cases}
\end{equation}
This yields the following displacement field,
\begin{widetext}
\begin{equation}
\begin{split}
u_x(z,\bm{\kp},\omega)&=\frac{-if^\mathrm{ext}_zk_x\left(e^{\ql z}(\kp^2+\qt^2)-2e^{\qt 
		z}\ql\qt\right)}
{\rho\left[\cl^2(\kp^2-\ql^2)(\kp^2+\qt^2)-2\ct^2\kp^2
	\left(\kp^2-2\qt\ql+\qt^2\right)\right]},\\
u_y(z,\bm{\kp},\omega)&=\frac{-if^\mathrm{ext}_zk_y\left(e^{\ql z}(\kp^2+\qt^2)-2e^{\qt 
		z}\ql\qt\right)}
{\rho\left[\cl^2(\kp^2-\ql^2)(\kp^2+\qt^2)-2\ct^2\kp^2
	\left(\kp^2-2\qt\ql+\qt^2\right)\right]},\\
u_z(z,\bm{\kp},\omega)&=\frac{f^\mathrm{ext}_z\ql\left(e^{\ql z}(\kp^2+\qt^2)-2e^{\qt 
		z}\kp^2\right)}
{\rho\left[\cl^2(\kp^2-\ql^2)(\kp^2+\qt^2)-2\ct^2\kp^2
	\left(\kp^2-2\qt\ql+\qt^2\right)\right]}.
\end{split}
\end{equation}
The second case is when the external force is parallel to the surface (in the $x$-direction),
\begin{equation}
\sigma_{iz}(z=0)=\begin{cases}
f^\mathrm{ext}_x\qquad&  \text{if}~i=x\\
0\qquad&  \text{otherwise}.
\end{cases}
\end{equation}
With this boundary condition, one arrives at
\begin{equation}
\begin{split}
u_{x}(z,\bm{\kp},\omega)&=\frac{f^\mathrm{ext}_x\left[\cl^2 e^{\qt z}\left(k_y^2+\qt^2\right)  
	\left(\kp^2-\ql^2\right)-2 \ct^2 \left[e^{\qt z} 
	\left(\kp^2\qt^2+k_y^2(\kp^2-2\ql\qt)\right)-k_x^2 \qt^2 e^{\ql 
		z}\right]\right]}
{\rho\ct^2\qt\left[\cl^2(\kp^2-\ql^2)(\kp^2+\qt^2)-2\ct^2\kp^2
	\left(\kp^2-2\qt\ql+\qt^2\right)\right]},\\
u_{y}(z,\bm{\kp},\omega)&=\frac{-k_x k_y f^\mathrm{ext}_x \left(\cl^2 
	e^{\qt z} \left(\kp^2-\ql^2\right)-2 \ct^2 
	\left(e^{\qt z} \left(\kp^2-2 \ql 
	\qt\right)+\qt^2 e^{\ql z}\right)\right)}
{\rho\ct^2\qt\left[\cl^2(\kp^2-\ql^2)(\kp^2+\qt^2)-2\ct^2\kp^2
	\left(\kp^2-2\qt\ql+\qt^2\right)\right]},\\
u_{z}(z,\bm{\kp},\omega)&=\frac{-i k_x f^\mathrm{ext}_x\left(\cl^2 e^{\qt z} 
	\left(\kp^2-\ql^2\right)+2 \ct^2 \left(\ql 
	\qt e^{\ql z}-\kp^2 e^{\qt 
		z}\right)\right)}
{\rho\ct^2\qt\left[\cl^2(\kp^2-\ql^2)(\kp^2+\qt^2)-2\ct^2\kp^2
	\left(\kp^2-2\qt\ql+\qt^2\right)\right]}.
\end{split}
\end{equation}
Due to the symmetry, the external force acting in the $y$-direction can be easily obtained by simply exchanging the indices $x\leftrightarrow y$.

Having obtained the solutions that solve~\cref{eq:ceqm2} with the boundary conditions~\cref{eq:st_bound}, we can find the Green's tensor from~\cref{eq:G},
\begin{equation}
\begin{aligned}
G_{xz}(z,\vkp,\omega)&=\frac{-ik_x\left(e^{\ql z}(\kp^2+\qt^2)-2e^{\qt 
		z}\ql\qt\right)}
{\rho\left[\cl^2(\kp^2-\ql^2)(\kp^2+\qt^2)-2\ct^2\kp^2
	\left(\kp^2-2\qt\ql+\qt^2\right)\right]},\\
G_{yz}(z,\vkp,\omega)&=\frac{-ik_y\left(e^{\ql z}(\kp^2+\qt^2)-2e^{\qt 
		z}\ql\qt\right)}
{\rho\left[\cl^2(\kp^2-\ql^2)(\kp^2+\qt^2)-2\ct^2\kp^2
	\left(\kp^2-2\qt\ql+\qt^2\right)\right]},\\
G_{zz}(z,\vkp,\omega)&=\frac{\ql\left(e^{\ql z}(\kp^2+\qt^2)-2e^{\qt 
		z}\kp^2\right)}
{\rho\left[\cl^2(\kp^2-\ql^2)(\kp^2+\qt^2)-2\ct^2\kp^2
	\left(\kp^2-2\qt\ql+\qt^2\right)\right]},\\
G_{xx}(z,\vkp,\omega)&=\frac{\cl^2 e^{\qt z}\left(k_y^2+\qt^2\right)  
	\left(\kp^2-\ql^2\right)-2 \ct^2 \left[e^{\qt z} 
	\left(\kp^2\qt^2+k_y^2(\kp^2-2\ql\qt)\right)-k_x^2 \qt^2 e^{\ql 
		z}\right]}
{\rho\ct^2\qt\left[\cl^2(\kp^2-\ql^2)(\kp^2+\qt^2)-2\ct^2\kp^2
	\left(\kp^2-2\qt\ql+\qt^2\right)\right]},\\
G_{yx}(z,\vkp,\omega)&=\frac{-k_x k_y \left(\cl^2 
	e^{\qt z} \left(\kp^2-\ql^2\right)-2 \ct^2 
	\left(e^{\qt z} \left(\kp^2-2 \ql 
	\qt\right)+\qt^2 e^{\ql z}\right)\right)}
{\rho\ct^2\qt\left[\cl^2(\kp^2-\ql^2)(\kp^2+\qt^2)-2\ct^2\kp^2
	\left(\kp^2-2\qt\ql+\qt^2\right)\right]}.
\label{eq:full_GT}
\end{aligned}
\end{equation}
\end{widetext}
Note that the Green's tensor have the following symmetry
\begin{equation}
G_{ij}(z,k_x,k_y,\omega)=G_{ji}(z,-k_x,-k_y,\omega).
\end{equation}
At $z\to0$ and $\Im{\ca}\to0$, $G_{ij}(\vkp,\omega)$ reduces to the expression reported in Refs.~\cite{persson85a,persson01a,volokitin06a}.

\section{Dynamical matrix of the viscoelastic simple cubic crystal}\label{ap:dispersion}
\begin{figure*}
	\centering
	\subfloat[]{\centering\label{fig:re_dispersion}
		\includegraphics[width=0.85\columnwidth]{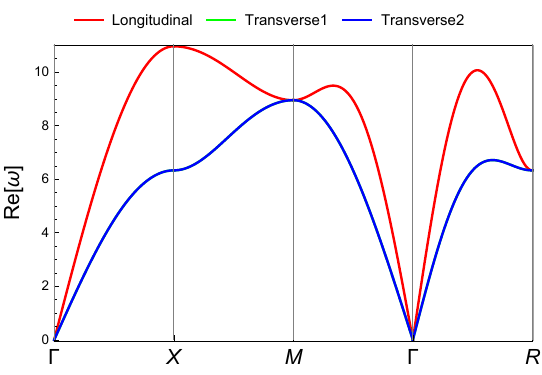}}
	\subfloat[]{\centering\label{fig:im_dispersion}
		\includegraphics[width=0.85\columnwidth]{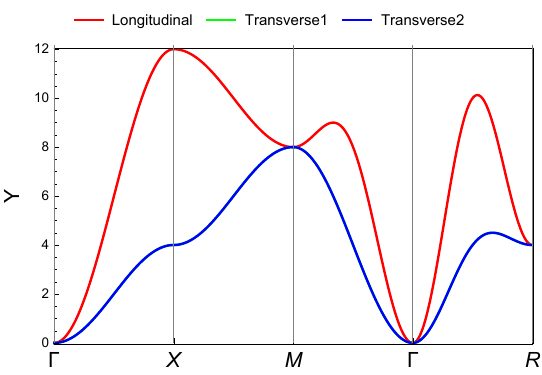}}
	\caption{The real (a) and imaginary (b) part of the dispersion relations for the longitudinal and transverse modes of 
		the simple cubic crystal with $M=a=1$, $\kappa_1=\kappa_2=10$, and $\gamma_1=\gamma_2=1$. Note that 
		there exist two transverse modes, and they are degenerate.}
	\label{fig:dispersion}
\end{figure*}
\begin{figure}
	\centering
	\includegraphics[width=\columnwidth]{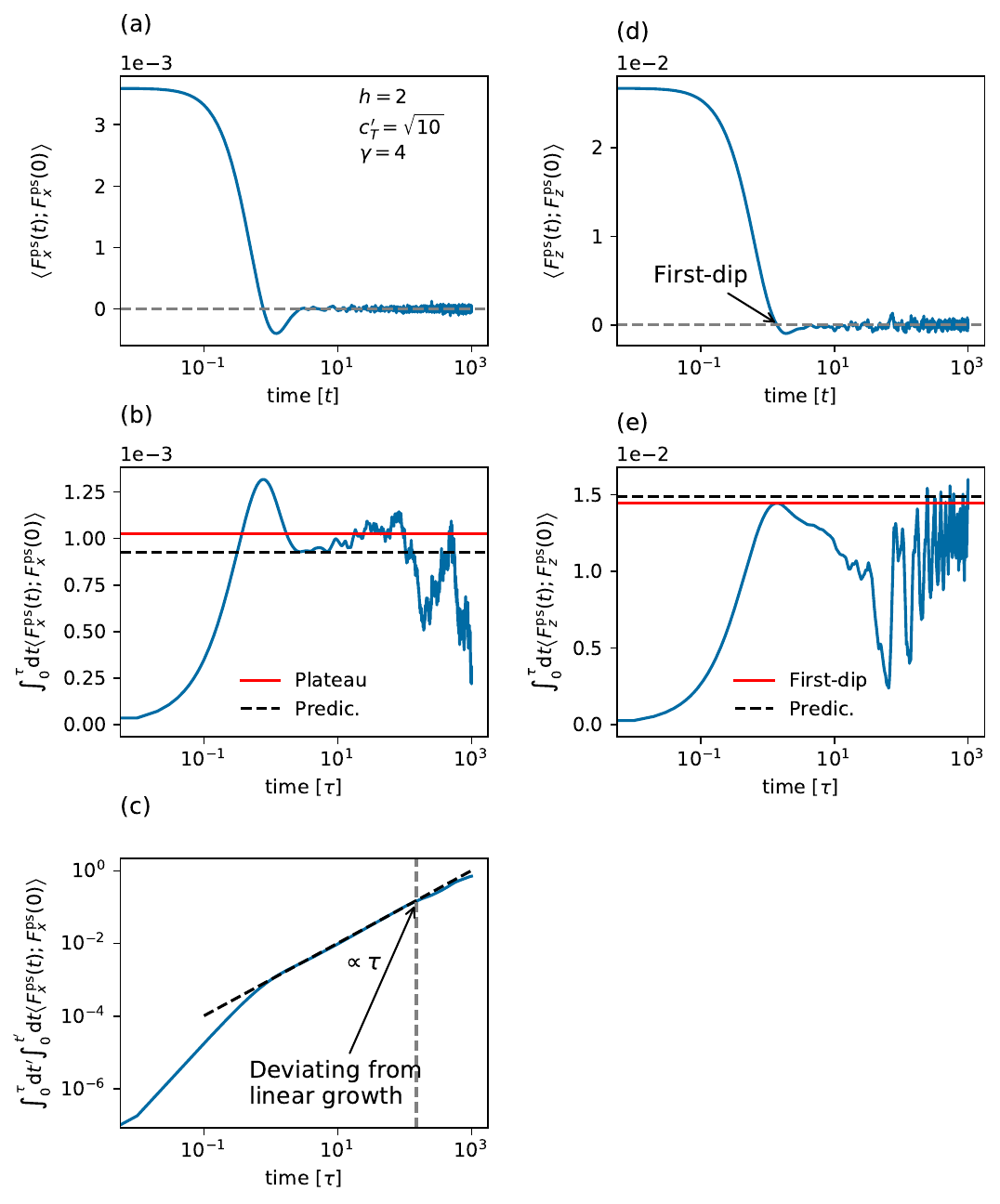}
	\caption{The probe-surface force covariance function of the (a) $xx$ and (d) $zz$-component, as a function of time, for parameters $h,c_T',\gamma$ as indicated. Panels (b) and (e) show the respective running value of the Green-Kubo integral, which ideally should converge to a plateau. Panel (c) shows the double running integral of the $xx$-component, where a regime of linear growth is clearly visible (extending to the vertical dashed line, marking the upper integration bound). The plateau in (b), represented by the red line, is determined from the slope of the double running integral. For the $zz$-component, the double integral method proved unsuccessful, and so here the first-dip method was used; panel (d) indicates the first-dip time, where the covariance function becomes zero for the first time. The red line in (e) indicates the corresponding estimate of the $zz$-component of the friction tensor. The black dashed lines in (b) and (e) denote the predictions of our theory.}
	\label{fig:plateau}
\end{figure}
The equation of motion of bulk solid atoms in Fourier space reads (the Einstein summation rule is assumed)
\begin{equation}\begin{aligned}
&\left[M\omega^2\delta_{ij} - D_{ij}(\bm{k},\omega)\right]U_{j}(\bm{k},\omega)=F^\mathrm{ran}_i(\bm{k},\omega)\\&\qquad\qquad\text{with}~i,j\in\{x,y,z\},
\end{aligned}\end{equation}
where $U_{j}(\bm{k},\omega)$ is the displacement of the solid atoms in the $j$-direction in Fourier space, $M$ the mass of each atom, and $D_{ij}(\bm{k},\omega)$ the dynamical matrix.
The Fourier transform is defined as
$U_i(\bm{R}_n,t)=\frac{1}{(2\pi)^4}\int_{-\pi/a}^{\pi/a}\dd^3{\bm{k}}\int_{-\infty}^\infty\dd{\omega}U_i(\bm{k},\omega) e^{i\left(\bm{k}\cdot\bm{R}_n-\omega t\right)}
$ with $a$ denoting the lattice constant.

Let us now consider the simple cubic crystal solid that we defined in~\cref{sec:theory}. The dynamical matrix of the viscoelastic crystal (the Einstein summation rule is not assumed) reads,
\begin{equation}\begin{aligned}
D_{ij}(\bm{k},\omega)=&4\delta_{ij}(\kappa_1-i\omega\gamma_1)\sin{\frac{ak_i}{2}}\sin{\frac{ak_j}{2} }\\
&+2(1-\delta_{ij})(\kappa_2-i\omega\gamma_2)\sin{ak_i}\sin{ak_j}\\
&+2\delta_{ij}(\kappa_2-i\omega\gamma_2)\\
&\quad\times\left[
2-\cos{ak_i}\left(\sum_{l}^{\{x,y,z\}}\cos{ak_l}-\cos{ak_i}\right)
\right],
\end{aligned}\end{equation}
from which one can calculate the dispersion relations at any point in the first Brillouin zone (see~\cref{fig:re_dispersion}).

Around $\bm{k}=0$, the dispersion relations are given by
\begin{equation}\begin{aligned}
\omega_\mathrm{L}(\bm{k})=&2\sqrt{\frac{\kappa_1+2\kappa_2
	-i\omega(\gamma_1+2\gamma_2)}{M}}\sin{\frac{ak}{2}},\\
\omega_\mathrm{T}(\bm{k})=&2\sqrt{\frac{\kappa_2-i\omega\gamma_2}{M}}\sin{\frac{ak}{2}}
\end{aligned}\end{equation}
with
\begin{equation*}
k = \abs{\bm{k}}.
\end{equation*}
Here, $\kappa_1$ and $\kappa_2$ ($\gamma_1$ and $\gamma_2$) are the spring (damping) coefficients among the nearest and the next nearest neighbors, respectively~\footnote{In fact, there exist two transverse modes. They are, however, degenerate around the $\Gamma$ point.}.

The speeds of sound of two modes are thus
\begin{equation}\begin{aligned}
\cl(\omega)=&a\sqrt{\frac{\kappa_1+2\kappa_2-i\omega(\gamma_1+2\gamma_2)}{M}},\\
\ct(\omega)=&a\sqrt{\frac{\kappa_2-i\omega\gamma_2}{M}}.
\label{eq:dspeed}
\end{aligned}\end{equation}

The fact that the solid has the dampers on the springs is manifested by the dynamical matrix being complex, the imaginary part of which accounts for the dissipative behavior of the system. To isolate the imaginary part and thus the dissipative behavior, let us define the following quantity~\cite{michel15a,landau86a},
\begin{equation}
\Upsilon_\alpha(\bm{k}) = -\Im{\frac{\omega_\alpha(\bm{k})}{M\omega}},
\end{equation}
where $\omega_\alpha$ is the eigenvalue of the $\alpha$ mode. As shown in~\cref{fig:im_dispersion}, $\Upsilon_\alpha(\bm{k})$ are quadratic in $k$ around the $\Gamma$ point,
\begin{equation}\begin{aligned}
\Upsilon_L(\bm{k}) =& \frac{a^2}{M}(\gamma_1+2\gamma_2)k^2\\
\Upsilon_T(\bm{k}) =& \frac{a^2}{M}\gamma_2k^2,
\end{aligned}\end{equation}
whose curvatures are the viscosity of the corresponding modes.

Notice that the simple cubic crystal can have three distinct modes depending on the spring and damping coefficients: one longitudinal and two transverse modes. With our choice of the spring and damping coefficients, however, the crystal behaves isotropic with respect to mechanical distortion since the two transverse modes are always degenerated. This complies with the continuum description where there only exist the transverse and longitudinal modes.

\section{Molecular Dynamics / Green-Kubo integral}\label{ap:simulation}
Solving the equations of motion, \cref{eq:deqm}, is done via standard molecular dynamics using LAMMPS~\cite{lammps} with some modifications to properly implement the stochastic dashpot springs. We emphasize that the spring forces are computed in the harmonic approximation, i.e.~retaining only linear terms in the particle displacements, to exclude any geometric anharmonicity~\cite{norell16a}. The principal output of these simulations is the trajectory, i.e.~a time series of the force on each atom, from which all quantities of interest can be computed. The integration time step $dt = 0.01 \tau$, where $\tau$ denotes the Lennard-Jones (LJ) time unit. Each simulation consists of 1.5 million integration steps; to allow for equilibration (temperature $T=0.15$), only the last million of these are used to collect the time series.

We simulate a simple cubic lattice consisting of $(X,Y,Z)=(100,100,102)$ unit cells, with lattice constant $a\equiv 1$, implying number density $n_\mathrm{A}=1$ in our LJ units. Periodic boundary conditions are applied in the lateral $x$ and $y$ directions, but not in the vertical $z$ direction. Each particle has unit mass, $M \equiv 1$, implying mass density $\rho=1$. Each atom is bonded to its nearest and next nearest neighbors via the dashpot springs of~\cref{eq:bond} (18 bonds per atom in total, except for atoms in the bottom and top vertical layer, which have less). We choose identical spring and damping coefficients for nearest and next nearest neighbor bonds: $\kappa_1 = \kappa_2 \equiv \kappa$ and $\gamma_1 = \gamma_2 \equiv \gamma$. The exception is for atoms in the bottom layer and those in the layer directly above; for these atoms, we set $\gamma=0$. The atoms in the bottom layer remain frozen at all times (to fixate the system in the vertical direction) while the atoms in the layer directly above have a standard Langevin thermostat applied to them, with damping parameter $\gamma_{\rm lan}=100 M/\tau$ (in order to regulate the temperature in cases where $\gamma$ of the dashpots is small, and to minimize phonon back-reflections~\cite{kantorovich08a_theory, kantorovich08b_theory, benassi10a_theory, benassi12a_theory}).

After each single time step, we calculate the pairwise forces between the fictitious probe and the surface atoms (i.e.~those in the top layer only) using~\cref{eq:pair_pot} with exponent $n=3$ at different heights $1 \leq h/a \leq 30$ (periodic images are not considered in this part of the calculation). That is, for given material properties ($\kappa, \gamma$), the pairwise forces at different heights are calculated from the same trajectory. Summing the pairwise force over all surface atoms yields the probe-surface force, which is then used to evaluate the Green-Kubo integral, as defined in~\cref{eq:ff_cor_bare}. Ideally, one expects the value of the integral to become constant at large times, i.e.~that a plateau be observed, but this condition is rarely met in practice~\cite{kiryl19a}. For this reason, the integral cannot be extended to arbitrary large times, but the introduction of an upper bound becomes necessary~\cite{chen10a, li98a}.

{\color{black}Since the $xx$-component of the probe-surface force covariance tensor decays in time, the running time integral yields a plateau for larger times. This implies that the \textit{double} running time integral grows linearly in time after the covariance vanishes (see~\cref{fig:plateau}(a)-(c)).} The upper bound can be found when the growth of the double running integral is no longer linear, at which point the data may be considered unreliable due to accumulation of statistical errors. The slope of the linear growth yields the plateau value as shown in~\cref{fig:plateau}(b). For the $zz$-component, unfortunately, a plateau is harder to identify with the double integral method. We believe this could be due to the back-reflection of phonons from the frozen bottom layer~\cite{kantorovich08a_theory, kantorovich08b_theory, vink19a_simulation, benassi10a_theory, benassi12a_theory}. To compute the $zz$-component of the friction tensor, we therefore resorted to the so-called first-dip method, which is a commonly used approximation to evaluate a Green-Kubo integral~\cite{li98a}. In this method, the upper bound of the Green-Kubo integral is taken to be where the covariance function becomes zero for the first time (see~\cref{fig:plateau}(d)-(e)).

\section{Friction tensor}\label{ap:friction_cal}
To find the expression that relates the friction tensor with the Green's tensor 
in Fourier space, we employ the following substitutions, starting 
from~\cref{eq:FDT_friction},
\begin{equation}\begin{aligned}
\Gamma_{ij}(h,\omega)
=&\frac{n_\mathrm{A}^2}{\omega}\Im\left\{ \iiiint_{-\infty}^{\infty}
\dd{x}\dd{y}\dd{x'}\dd{y'}\right.\\
&\times\left.f_{i,k}G_{kl}(x,y,x',y',\omega)f'_{j,l} \right\}\\
=&\frac{1}{\omega}\frac{\na^2}{(2\pi)^2}\Im\left\{\iiiint_{-\infty}^{\infty}
\dd{x}\dd{y}\dd{x'}\dd{y'}f_{i,k}f'_{j,l}\right.\\
&\times\left.\iint_{-\infty}^{\infty}\dd{k_x}\dd{k_y}G_{kl}(k_x,k_y,\omega)e^{i(k_x
(x-x')+k_y(y-y'))}\right\}
\label{eq:cov1}
\end{aligned}\end{equation}
with $\partial_jf_i = f_{i,j}$ and $f' = f(\bm{d}')$.
Performing the Fourier transforms on the forces, we arrive at
\begin{equation}\begin{aligned}
\Gamma_{ij}(h,\omega)
=&\frac{1}{\omega}\frac{n_\mathrm{A}^2}{(2\pi)^2}\Im\left\{\iint_{-\infty}^{\infty}
\dd{k_x}\dd{k_y}\right.\\
&\left.\times 
f_{i,k}^*(k_x,k_y,h)G_{kl}(k_x,k_y,\omega)f_{j,l}(k_x,k_y,h)\right\}\\
=&\frac{1}{\omega}\frac{1}{(2\pi)^2}\Im{\iint_{-\infty}^{\infty}\dd{k_x}\dd{k_y}
H_{ij}(k_x,k_y,\omega)}\\
=&\frac{1}{\omega}\frac{1}{(2\pi)^2}\Im{\int_{0}^{\infty}\dd{\kp}\int_{0}^{2\pi}
\dd{\theta} \kp H_{ij}(\bm{\kp},\omega)},
\label{eq:ff_cor_general}
\end{aligned}\end{equation}
where
\begin{equation}\begin{aligned}
H_{ij}(\vkp,h,\omega)\equiv n_\mathrm{A}^2
f_{i,k}^*(\vkp,h)G_{kl}(\vkp,\omega)f_{j,l}(\vkp,h)
\label{eq:ff_cor_integrand}
\end{aligned}\end{equation}
with $f_{i,k}^*(\vkp,h)$ being the complex conjugate of $f_{i,k}(\vkp,h)$.
\begin{figure}
\centering
\includegraphics[width=0.7\columnwidth]{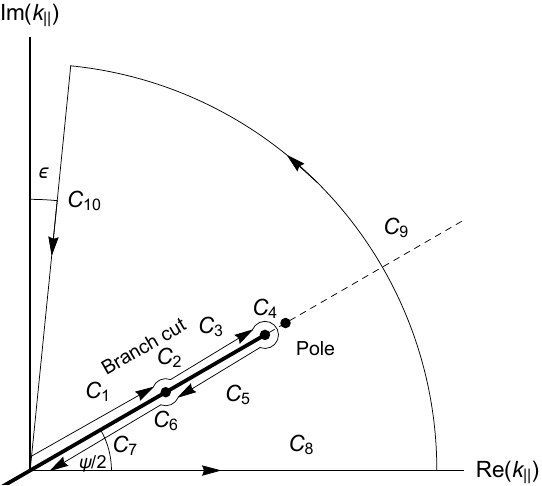}
\caption{The branch points, the corresponding branch cuts, and the pole on the 
complex 
plane of $\kp$, when $\cl(\omega)=\sqrt{3}\ct(\omega)$. The branch points and the pole are aligned with the slop defined by $\psi/2$.
The contour path is indicated by the arrows, and the radius of the arc is infinitely large. The small angle $\epsilon$ is 
introduced to ensure the convergence of the integration.}
\label{fig:contour1}
\end{figure}

The problem, therefore, comes down to identifying $H_{ij}(\vkp,h,\omega)$ and performing the integration analytically or numerically.

For $n=3$ in~\cref{eq:pair_int}, 
the pairwise potential in Fourier space reads
\begin{equation}
V(\vkp,h)=\frac{2\pi n_\mathrm{A}\alpha e^{-\kp h}}{h}.
\label{eq:pair_pot}
\end{equation}

Since the Green's tensor in~\cref{eq:full_GT} bares poles, 
~\cref{eq:ff_cor_general} is an improper integral. A value to the integration 
has to be assigned by means of the residue theorem~\cite{arfken99a,dhont96a}.

The contour integral to assign a proper value to~\cref{eq:ff_cor_general} is shown in~\cref{fig:contour1}. The poles are located at, assuming $b=\cl/\ct=\sqrt{3}$,
\begin{equation}
k_c=\pm\frac{\sqrt{3+\sqrt{3}}}{2}\frac{\omega}{\abs{\ct}}e^{\frac{i\psi}{2}}.
\end{equation}
The contour integral consists of the following parts
\begin{equation}\begin{aligned}
&\oint\dd{\kp} H^{\theta}_{ij}(\kp,h,\omega) = \int_{C_1+C_2+\cdots+C_{10}}\dd{\kp} H^{\theta}_{ij}(\kp,h,\omega),
\end{aligned}\end{equation}
where
\begin{equation}\begin{aligned}
H^{\theta}_{ij}(\kp,h,\omega)&=\int_{0}^{2\pi}\dd{\theta}\kp 
H_{ij}(\vkp,h,\omega),
\end{aligned}\end{equation}
which is a diagonal matrix. Notice that $H^{\theta}_{ij}$ is a function of $\kp$ rather than $\vkp$.

Integrating around the arcs $C_2$,$C_4$, and $C_9$ does not contribute to the finial result. The right hand side of the above equation is given by the residue,
\begin{equation}\begin{aligned}
\oint\dd{\kp} H^{\theta}_{ij}(\kp,h,\omega) = 2\pi i\Res{
H^{\theta}_{ij}(\kp,h,\omega)}_{\text{at}~\kp=k_c}.
\end{aligned}\end{equation}
Re-arranging the terms such that the integration running from $0$ to $\infty$, we can find the value of the force covariance.

Notice that although integrating around the branch cut and the pole can be analytically done, integrating along $C_{10}$ seems to be not possible. We thus make the the following simplification; we expand $H^{\theta}_{ij}(\kp,h,\omega)$ at small $\omega$ first, and then performing the integration over $\kp$-space. The consequence of the frequency expansion is that our final results only hold for the first leading terms in $\omega$.

Let us perform the variable change 
$\kp=\frac{\omega}{\abs{\ct}}e^{i\frac{\psi}{2}}p$ in $H^{\theta}_{ij}(\kp,h,\omega)$,
\begin{widetext}
\begin{equation}
\resizebox{0.9\textwidth}{!}{$\begin{aligned}
	H^{\theta}_{xx}(p,h,\omega)=&-\frac{n_\mathrm{A}^2 4\pi^3\alpha^2
		e^{-2ph\frac{\omega}{\abs{\ct}}e^{i\frac{\psi}{2}}}p^3\omega^2}{h^4\rho\abs{\ct}^4
		\ct^2\left(3\left(2p^2-1\right)^2-4\sqrt{3}p^2\sqrt{p^2-1}\sqrt{3p^2-1}\right)}
	\left[\sqrt{3}\abs{\ct}^2e^{i\psi}\sqrt{\left(3 p^2-1\right)}\right.\\
	&\left.+2 p h \abs{\ct}\omega e^{i\frac{3\psi}{2}}\left(-3p\left(2 
	p^2-1\right)+\sqrt{3}\sqrt{\left(3 p^2-1\right)}+2\sqrt{3}p 
	\sqrt{\left(p^2-1\right)} \sqrt{\left(3 p^2-1\right)}\right)
	\right.\\
	&\left.+p^2h^2\omega^2e^{i2 \psi}\left(-6p \left(2p^2-1\right)+4 \sqrt{3}p 
	\sqrt{\left(p^2-1\right)} \sqrt{\left(3p^2-1\right)}+\left(3 
	\sqrt{\left(p^2-1\right)}+\sqrt{3} \sqrt{\left(3 
		p^2-1\right)}\right)\right)\right].
\end{aligned}$}
\end{equation}
\begin{equation}
\resizebox{0.9\textwidth}{!}{$\begin{aligned}
	H^{\theta}_{zz}(p,h,\omega)=&-\frac{n_\mathrm{A}^2 8\pi^3\alpha^2
		e^{-2ph\frac{\omega}{\abs{\ct}}e^{i\frac{\psi}{2}}}p}{h^6\rho\abs{\ct}^4
		\ct^2\left(3\left(2p^2-1\right)^2-4\sqrt{3}p^2\sqrt{p^2-1}\sqrt{3p^2-1}\right)}
	\left[4\sqrt{3} \abs{\ct}^4 \sqrt{\left(3 
		p^2-1\right)}\right.\\
	&\left.+4 p h \abs{\ct}^3 \omega e^{i\frac{\psi}{2}} 
	\left(-3p\left(2p^2-1\right) +2 
	\sqrt{3}\sqrt{\left(3 p^2-1\right)}+2 
	\sqrt{3} p \sqrt{\left(p^2-1\right)} \sqrt{\left(3 p^2-1\right)}\right)
	\right.\\
	&\left.+p^2 h^2\abs{\ct}^2\omega^2e^{i\psi } \left(-24 p 
	\left(2 p^2-1\right)+16 \sqrt{3} p \sqrt{\left(p^2-1\right)}\sqrt{\left(3 
		p^2-1\right)}+\left(3\sqrt{\left(p^2-1\right)}+8\sqrt{3}\sqrt{\left(3
		p^2-1\right)}\right)\right)\right.\\
	&\left.+2 p^3 h^3 \abs{\ct}\omega^3 e^{i \frac{3\psi}{2} }\left(-9 p \left(2 
	p^2-1\right)+6 \sqrt{3} p \sqrt{\left(p^2-1\right)} 
	\sqrt{\left(3 p^2-1\right)}+\left(3 
	\sqrt{\left(p^2-1\right)}+2 \sqrt{3} \sqrt{\left(3 p^2-1\right) 
	}\right)\right)\right.\\
	&\left.+p^4 h^4 \omega^4 e^{4 i \psi} 
	\left(-6 p \left(2 p^2-1\right) 
	+4 \sqrt{3} p \sqrt{\left(p^2-1\right)} \sqrt{\left(3 
		p^2-1\right)}+\left(3 
	\sqrt{\left(p^2-1\right)}+\sqrt{3} \sqrt{\left(3 p^2-1\right) 
	}\right)\right)\right].
\end{aligned}$}
\label{eq:full_H}
\end{equation}
Expanding $e^{-2ph\frac{\omega}{\abs{\ct}}}$ and $\psi(\omega)$ at small $\omega$ leaves us with
\begin{equation}
H^{\theta}_{xx}(p,h,\omega)=-\frac{n_\mathrm{A}^2 4\sqrt{3}\pi^3\alpha^2p^3\omega^2\abs{\ct}^2
\sqrt{\left(3p^2-1\right)}}{h^4\rho\abs{\ct}^4
\ct^2\left(3\left(2p^2-1\right)^2-4\sqrt{3}p^2\sqrt{p^2-1}\sqrt{3p^2-1}\right)}+
\order{\omega^3},
\end{equation}
\begin{equation}
H^{\theta}_{zz}(p,h,\omega)=-\frac{n_\mathrm{A}^2 32\sqrt{3}\pi^3\alpha^2p  \abs{\ct}^4 
\sqrt{\left(3 p^2-1\right)}}{h^6\rho\abs{\ct}^4
\ct^2\left(3\left(2p^2-1\right)^2-4\sqrt{3}p^2\sqrt{p^2-1}\sqrt{3p^2-1}\right)}+
\order{\omega}.
\end{equation}
\end{widetext}

\begin{figure}
\centering
\includegraphics[scale=0.7]{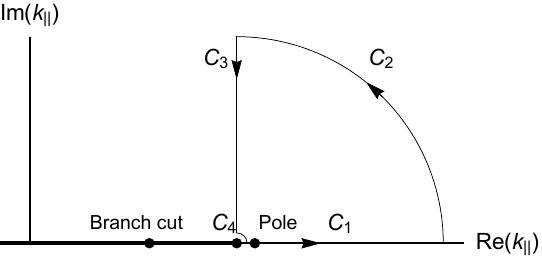}
\caption{The branch points and the pole on the 
complex 
plane of $\kp$ after truncating $\psi(\omega)$ at the leading order of $\omega$, assuming 
$\cl(\omega)=\sqrt{3}\ct(\omega)$.}
\label{fig:contour_zero}
\end{figure}

As shown in~\cref{fig:contour_zero}, the branch points and the pole are now on the real axis.
The integration from $0$ to $\Re{k_{b2}}$ thus has to be dissected 
as
\begin{equation}\begin{aligned}
&\int_0^{\Re{k_{b2}}}\dd{p}
\frac{\omega}{\abs{\ct}}H^\theta_{ij}(p,h,\omega)
\\&\quad=\int_0^{\sqrt{\frac{1}{3}}}\dd{p}
\frac{\omega}{\abs{\ct}}H^\theta_{ij}(p,h,\omega)
+\int_{\sqrt{\frac{1}{3}}}^{1}\dd{p}
\frac{\omega}{\abs{\ct}}H^\theta_{ij}(p,h,\omega).
\end{aligned}\end{equation}
This yields for the $xx$ entry
\begin{equation}\begin{aligned}
\int_0^{\sqrt{\frac{1}{3}}}\dd{p}
\frac{\omega}{\abs{\ct}}H^\theta_{xx}(p,h,\omega)=&\frac{1.9i n_\mathrm{A}^2\alpha ^2 
\omega^3}{h^4 \rho\ct^2\abs{\ct}^3 }+\order{\omega^4}\\
\int_{\sqrt{\frac{1}{3}}}^{1}\dd{p}
\frac{\omega}{\abs{\ct}}H^\theta_{xx}(p,h,\omega)=&\frac{(6.7+16.6 
i)n_\mathrm{A}^2\alpha ^2 
\omega ^3}{h^4\rho\ct^2\abs{\ct}^3}+\order{\omega^4},
\end{aligned}\end{equation}
and for the $zz$ entry
\begin{equation}\begin{aligned}
\int_0^{\sqrt{\frac{1}{3}}}\dd{p}
\frac{\omega}{\abs{\ct}}H^\theta_{zz}(p,h,\omega)=&\frac{93.9i n_\mathrm{A}^2\alpha 
^2\omega}{h^6 
\rho\ct^2\abs{\ct} }+\order{\omega^2}\\
\int_{\sqrt{\frac{1}{3}}}^{1}\dd{p}
\frac{\omega}{\abs{\ct}}H^\theta_{zz}(p,h,\omega)=&\frac{(67.2+207.5i) n_\mathrm{A}^2\alpha ^2\omega}{h^6 \rho\ct^2\abs{\ct} }+\order{\omega^2}.
\end{aligned}\end{equation}

The remaining integral can be evaluated using a Cauchy principal value 
\begin{equation}
\begin{split}
&\oint\dd{\kp}H^\theta_{ij}(\kp,h,\omega) = \int_{C_1+C_2+C_3+C_4}\dd{\kp}H^\theta_{ij}(\kp,h,\omega)\\
&\qquad=\pi i\Res{H^\theta_{ij}(\kp,h,\omega)}_{\text{at}~\kp=k_c}.
\end{split}
\end{equation}

Again, integrating around the arcs $C_2$ and $C_4$ does not contribute. As a result, one arrives at
\begin{equation}
\begin{split}
&\int_{\Re{k_{b2}}}^{\infty}\dd{\kp}H^\theta_{ij}(\kp,h,\omega) =- \int_{i\infty+\Re{k_{b2}}}^{\Re{k_{b2}}}\dd{\kp}H^\theta_{ij}(\kp,h,\omega)\\
&\qquad+\pi i\Res{H^\theta_{ij}(\kp,h,\omega)}_{\text{at}~\kp=k_c}.
\end{split}
\end{equation}

The Cauchy principal values are
\begin{equation}\begin{aligned}
&\pi i 
\Res{H^\theta_{xx}(\kp,h,\omega)}_{\text{at}~\kp=k_c}\\
&\qquad\qquad\qquad=
\frac{92.0in_\mathrm{A}^2\alpha^2\omega^3}{h^4\rho\ct^2\abs{\ct}^3}+\order{\omega^4}\\
&\pi i 
\Res{H^\theta_{zz}(\kp,h,\omega)}_{\text{at}~\kp=k_c}\\
&\qquad\qquad\qquad=
\frac{622.1in_\mathrm{A}^2\alpha^2\omega}{h^6\rho\ct^2\abs{\ct}}+\order{\omega^2}.
\end{aligned}\end{equation}

\begin{figure}
\centering
\includegraphics[width=1\columnwidth]{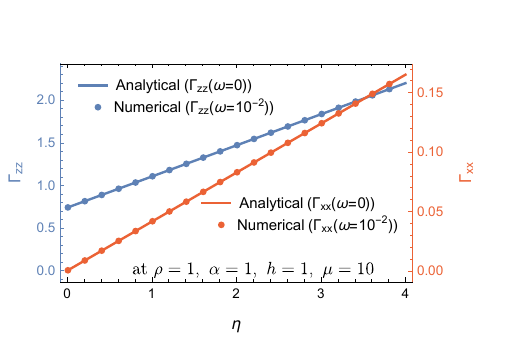}
\caption{The analytic expression of $\Gamma_{xx}$ and $\Gamma_{zz}$ compared against the numerical results by directly integrating~\cref{eq:ff_cor_general} with~\cref{eq:pair_pot}.}
\label{fig:numeric_esti}
\end{figure}

The integration along the imaginary axis can analytically be done after 
expanding $H^{\theta}_{ij}(\kp,h,\omega)$ at small $\omega$. This part is thus associated with the 
damping of phonons.
\begin{equation}\begin{aligned}
H^{\theta}_{xx}(\kp,h,\omega)=&\frac{\pi^3n_\mathrm{A}^2\alpha^2 
e^{-2h\kp}}{\ct^2\rho}\left(\frac{4\kp^4}{h^2}+\frac{4 \kp^3}{h^3}+\frac{3\kp^2 
}{h^4
}\right)+\order{\omega^2}\\
H^{\theta}_{zz}(\kp,h,\omega)=&\frac{\pi^3n_\mathrm{A}^2\alpha^2 
e^{-2h\kp}}{\ct^2\rho}\left(
\frac{8\kp^4}{h^2}
+\frac{24\kp^3}{h^3}
+\frac{38\kp^2}{h^4}\right.\\&\qquad\left.
+\frac{40\kp}{h^5}
+\frac{24}{h^6}\right)
+\order{\omega^2}.
\end{aligned}\end{equation}
Noting $\lim_{\omega\to0}k_{b2}= 0$, one arrives at
\begin{equation}\begin{aligned}
\lim_{\epsilon\to0}\int_{i\infty+\epsilon}^{0}\dd{\kp}
H^\theta_{xx}(\kp,h,\omega)&=-\frac{21\pi^3n_\mathrm{A}^2\alpha^2}{4h^7\rho\ct^2}+\order{\omega^2}\\
\lim_{\epsilon\to0}\int_{i\infty+\epsilon}^{0}\dd{\kp}
H^\theta_{zz}(\kp,h,\omega)&=-\frac{93\pi^3n_\mathrm{A}^2\alpha^2}{2h^7\rho\ct^2}+\order{\omega^2}
\label{apeq:img_int}
\end{aligned}\end{equation}
with $\epsilon$ to ensure the convergence of the integration.

Putting all the terms together we arrive at
\begin{equation}\begin{aligned}
\int_0^{\infty}\dd{\kp}H^\theta_{xx}(\kp,h,\omega)=&\frac{n_\mathrm{A}^2\alpha^2}
{\rho\ct^2}\left(\frac{(6.7+110.5i)\omega^3}{\abs{\ct}^3h^4}
+\frac{21\pi^3}{4h^7}\right)\\&+\order{\omega^4}\\
\int_0^{\infty}\dd{\kp}H^\theta_{zz}(\kp,h,\omega)=&\frac{n_\mathrm{A}^2\alpha^2}
{\rho\ct^2}\left(\frac{(67.2+923.5i)\omega}{\abs{\ct}h^6}
+\frac{93\pi^3}{2h^7}\right)\\&+\order{\omega^2}.
\end{aligned}\end{equation}
Plugging these results in~\cref{eq:ff_cor_general} yields the friction tensor,
\begin{equation}\begin{aligned}
\Gamma_{xx}(h,\omega)=&\frac{\alpha^2n_\mathrm{A}^2}{4\pi^2\rho\ct'^3}\left(
\frac{110.5\omega^2}{\ct'^2h^4}+\frac{325.6}{h^7}\frac{\ct''}{\omega}+\cdots\right)\\
\Gamma_{zz}(h,\omega)=&\frac{\alpha^2n_\mathrm{A}^2}{4\pi^2\rho\ct'^3}\left(
\frac{923.5}{h^6}+\frac{2883.6}{h^7}\frac{\ct''}{\omega}+\cdots\right),
\label{eq:f_covariance}
\end{aligned}\end{equation}
where
\begin{equation}\begin{aligned}
\ct'(\omega)&=\sqrt{\frac{\mu}{\rho}}\left(1+\frac{\eta^2\omega^2}{8\mu^2}+\order{\omega^4}\right)\\
\ct''(\omega)&=\ct'(0)\left(\frac{\eta\omega}{2\mu}-\frac{\eta^3\omega^3}{16\mu^3}+\order{\omega^5}\right)\\
&=\frac{\eta\omega}{2\rho\ct'(0)}-\frac{\eta^3\omega^3}{16\rho^3\ct'^5(0)}+\order{\omega^5}.
\end{aligned}\end{equation}
Due to the symmetry, $\Gamma_{yy}$ is identical to $\Gamma_{xx}$.

Replacing the inverse powers of height by the probe-surface force, one finds
\begin{equation}\begin{aligned}
\Gamma_{xx}(h,\omega)=&\frac{1}{4\pi^2\rho\ct'^3}\left(
\frac{2.8\omega^2}{\ct'^2}f_{\mathrm{ps}}^2(h)+\frac{2.06\ct''}{\omega}\frac{f_{\mathrm{ps}}'^2(h)}{h}+\cdots
\right)\\
\Gamma_{zz}(h,\omega)=&\frac{1}{4\pi^2\rho\ct'^3}\left(
5.9 f_{\mathrm{ps}}'^2(h)+\frac{18.3\ct''}{\omega}\frac{f_{\mathrm{ps}}'^2(h)}{h} +\cdots
\right).
\label{eq:ff_covariance}
\end{aligned}\end{equation}
The correctness of the calculated is double checked by numerically estimating 
the integral in~\cref{eq:ff_cor_general} as shown in~\cref{fig:numeric_esti}.

\section{On the ratio of the speeds of sound}
\label{ap:cratio}
In obtaining~\cref{eq:dfriction_results}, we assume $b=\cl/\ct=\sqrt{3}$. Here we explain the rationale behind the ratio. It is experimentally known that the ratio of the bulk and shear moduli is roughly 2, i.e., $K/\mu\approx 2$ in xeon~\cite{gornall71a} and krypton~\cite{petert73a}. Consequently, the ratio of the speeds of sound becomes $\cl/\ct =\sqrt{10/3}$ at $\omega=0$. We approximate this ratio to $\cl/\ct =\sqrt{3}$ at $\omega=0$, which can be achieved when $\kappa_1=\kappa_2$.

We could not, however, find an experimental report on the ratio of the imaginary parts of the speeds of sound. We thus assume that the same ratio translates to the imaginary parts, making it $\cl/\ct =\sqrt{3}$ for all $\omega$. This can be found when $\gamma_1=\gamma_2$.

\section{Highly localized force}\label{ap:local}

Performing the integral defined in~\cref{eq:ff_cor_general} with the potential in~\cref{eq:short_int}, one realizes that the elastic contribution yields the same results as in~\cref{eq:ff_covariance} at the given ratio of the speeds of sound, $b=\sqrt{3}$. The viscous contributions is however now changed, which is obtained from the integral along the imaginary axis,
\begin{equation}\begin{aligned}
&\lim_{\epsilon\to0}\int_{i\infty+\epsilon}^{0}\dd{\kp}
H^\theta_{xx}(\kp,h,\omega)\\
&=\frac{n_\mathrm{A}^2\pi^3}{8 l\ct^2\rho}
\left(9\sqrt{2\pi}V^2(h)+8l V(h)V'(h) + 3l^2\sqrt{2\pi}V'^2(h) \right)\\
&\lim_{\epsilon\to0}\int_{i\infty+\epsilon}^{0}\dd{\kp}
H^\theta_{zz}(\kp,h,\omega)\\
&=\frac{n_\mathrm{A}^2l\pi^3}{4\ct^2\rho}
\left(3\sqrt{2\pi}V'^2(h)+4l V'(h)V''(h) + 3l^2\sqrt{2\pi}V''^2(h) \right).
\end{aligned}\end{equation}
This leads to the friction tensor in~\cref{eq:friction_range}

\bibliography{bibtex/miru}
\bibliographystyle{apsrev4-2}
\end{document}